\documentclass[a4paper,showpacs,showkeys]{revtex4}
\usepackage[centertags]{amsmath}
\usepackage{amssymb}
\usepackage{latexsym}
\usepackage[]{graphicx}
\usepackage{subfigure}
\usepackage[]{hyperref}
\begin{document}
\title{Single-particle spectral function for the classical one-component plasma}
\author{C. Fortmann}
\email{carsten.fortmann@uni-rostock.de}
\homepage{http://everest.mpg.uni-rostock.de/~carsten}
\affiliation{Institut f\"ur Physik, Universit\"at Rostock, 18051 Rostock, Germany}
\begin{abstract}
  \noindent
  The spectral function for an electron
  one-component plasma is calculated self-consistently
  using the $GW^{(0)}$ approximation for the single-particle self-energy.
  In this way, correlation effects which go beyond the mean-field description of the plasma are
  contained, i.e. the collisional damping of single-particle states, 
  the dynamical screening of the interaction and the appearance of collective plasma modes. 
  Secondly, a novel non-perturbative analytic solution for the on-shell $GW^{(0)}$ self-energy as a function of momentum is presented.
  It reproduces the numerical data for the spectral function with a
  relative error of less than 10\% in the regime where the Debye screening parameter is smaller than the inverse Bohr radius,
  $\kappa<1\,a_\mathrm{B}^{-1}$. In the limit of low density, the non-perturbative self-energy behaves as
  $n^{1/4}$, whereas a perturbation expansion leads to the unphysical result of a density independent self-energy [W. Fennel and H. P. Wilfer,
  Ann. Phys. Lpz. \textbf{32}, 265 (1974)].
  The derived expression will greatly facilitate the calculation of observables in correlated plasmas
  (transport properties, equation of state) that need the spectral function as an input quantity. 
  This is demonstrated for the shift of the chemical potential,  which is 
  computed from the analytical formulae and compared to the 
  $GW^{(0)}$-result. At a plasma temperature of 100 eV and
  densities below $10^{21}\, \mathrm{cm^{-3}}$,
  both approaches deviate less than 10\% from each other.

\end{abstract}


\maketitle

\section{Introduction\label{sec:intro}}

The many-particle Green function approach \cite{KadanoffBaym:Book} 
allows for a systematic study of macroscopic properties of correlated systems. Green functions know
a long history of applications in solid state theory \cite{Mahan:Book}, nuclear \cite{FetterWalecka_1971}, and hadron physics
\cite{Hoell:2006AIPC..857...46H}, and also in the theory of strongly coupled plasmas \cite{Wierling_AdvPlasPhysRes_2002}.
In the latter case, optical and dielectric properties \cite{Reinholz:AnalDePhys06,Fortmann:CPP47_2007} have been studied using the Green function
approach, as well as transport properties like conductivity \cite{Reinholz_PRE.69.066412_2008} and 
stopping power \cite{Gericke:PhysLettA222_241_1996,KraeftStrege_PhysicaA149_313_1988}, and the equation of state \cite{Vorberger_PRE69_046407_2004}. Modifications of these
quantities due to the interaction among the constituents can be accessed, starting from a common starting point, namely the Hamiltonian of the
system. 

The key quantity to electronic properties in a correlated many-body environment is the electron spectral function $A(\mathbf{p},\omega)$, i.e. the
probability density to find an electron at energy (frequency) $\omega$ for a given momentum $\mathbf{p}$. It is related to the retarded electron self-energy
$\Sigma(\mathbf{p},\omega+i\delta), \delta>0$ via Dyson's equation
\begin{equation}
  A(\mathbf{p},\omega)=\frac{-2\mathrm{Im}\,\Sigma(\mathbf{p},\omega+i\delta)}{\left[ \omega-\varepsilon_\mathbf{p}
  -\mathrm{Re}\,\Sigma(\mathbf{p},\omega)\right]^2+\left[ \mathrm{Im}\,\Sigma(\mathbf{p},\omega+i\delta) \right]^2}~.
  \label{eqn:Dyson}
\end{equation}
Here, the single-particle energy
\begin{equation}
  \varepsilon_\mathbf{p}=p^2-\mu_\mathrm{e}
  \label{eqn:singleparticleenergy}
\end{equation}
has been introduced, $\mu_\mathrm{e}$ is the electron chemical potential.
Note that here and throughout the paper, the Rydberg system of units is used, where 
$\hbar=1$, $m_\mathrm{e}=1/2$, and $e^2/4\pi\epsilon_0=2$. Furthermore, the Boltzmann constant $k_\mathrm{B}$ is set
equal to 1, i.e. temperatures are measured in units of energy.

The self-energy describes the influence of correlations on the behaviour of the electrons. A finite, frequency dependent self-energy leads to a
finite life-time of single-particle states and a modification of the single-particle dispersion relation. 
Hence, the calculation
of the electron self-energy is the central task if one wants to determine electronic properties, e.g. those mentioned above.

The Hartree-Fock approximation \cite{KraeftKrempEbelingRoepke1986} represents the lowest order 
in a perturbative expansion of the self-energy in terms of the interaction potential \cite{EbelingKraeftKremp_1972}. 
Being a mean-field approximation, effects due to correlations in the system cannot be described. Examples are the appearance of 
collective modes, the energy transfer during particle collisions, and the quasi-particle damping. 
The next order term is the Born approximation, where binary collision are taken into
account via a bare Coulomb potential. 
However, the Born approximation leads to a divergent integral, due to the long-range Coulomb interaction. 
Therefore, the perturbation expansion of the
self-energy has to  be replaced by a non-perturbative approach, 
accounting for the dynamical screening of the interparticle potential. 

A non-perturbative approach to the many-particle problem is given by the theory of Dyson 
\cite{Dyson_PhysRev.75.1736} and Schwinger \cite{Schwinger_PNAS37_452_1951,Schwinger_PNAS37_455_1951} generalized to finite temperature and finite
density \cite{RobertsSchmidt_ProgNucPartPhys45_2000}. An excellent introduction to Dyson-Schwinger equations can also be found in
\cite{Hoell:2006AIPC..857...46H}.
The Dyson-Schwinger equation for the self-energy $\Sigma$ contains the full Green function $G$,
the screened interaction $W$ and the proper vertex $\Gamma$. Since each of these functions obeys a different Dyson-Schwinger equation itself,
involving higher order correlation functions, the Dyson-Schwinger approach leads to a hierarchy of coupled integro-differential equations. In
order to provide soluble equations, this hierarchy has to be closed at some level, i.e. correlation functions of a certain order either have to be
parametrized or neglected. 

One such closure of the Dyson-Schwinger hierarchy consists in neglecting the vertex, i.e. the three-point function, and
considering only the particle propagators and their respective self-energies, i.e. two-point functions. One arrives at the so-called $GW$-approximation, introduced in
solid-state physics by 
Hedin \cite{Hedin:PhysRev139_1965}. Hedin was led by the idea, to include correlations in the self-energy by replacing the Coulomb potential in
the Hartree-Fock self-energy by
the dynamically screened interaction $W$. 
In this way, one obtains a self-consistent, closed set of equations for the self-energy, the polarization function $\Pi$, the Green function and
the screened interaction.

It can be shown \cite{Barth_PRB72_235109}, that the $GW$ approximation belongs to the
so-called $\Phi$-derivable approximations \cite{BaymKadanoff_PhysRev.124.287,Baym_PhysRev.127.1391}. As such, it
leads to energy, momentum, and particle number conserving expressions for higher order
correlation functions. It was successfully applied in virtually all branches of solid state physics. An overview on theoretical foundations and applications of 
the $GW$ approximation can be found in the review articles
\cite{Aryasetiawan:RepProgPhys61_1998,Onida:RevModPhys.74.601,Mahan:CommCondMatPhys16_1994}.

The drawback of the $GW$ approximation is, that the Ward-Takahashi identities are violated. Ward-Takahashi identities 
provide an exact relation between the vertex function
$\Gamma$, i.e. the effective electron-photon coupling in the medium, and the self-energy and follow from the Dyson-Schwinger equations.
They reflect the gauge invariance of the theory. In $GW$, they are violated
simply because corrections to the vertex beyond zero order are neglected altogether. 
This issue touches on a fundamental problem in 
many-body theory and field theory, namely the question how to preserve gauge invariance in an effective, i.e. approximative theory, without
violating basic conservation laws. A detailed analysis of this question with application to nuclear physics is presented in a series of papers by van Hees and Knoll
\cite{vHeesKnoll_PhysRevD.65.025010,PhysRevD.65.105005,PhysRevD.66.025028}. 

Approximations for the self-energy, that also contain the vertex are often referred to as $GW\Gamma$ approximations. An example can be found in
Ref.~\cite{Takada:PhysRevLett.87.226402}, where the spectral function of electrons in aluminum is calculated using a parametrized vertex function.
An interesting result obtained in that work is that vertex corrections and self-energy corrections 
entering the polarization function, largely cancel. This can be understood as a consequence of Ward-Takahashi identities.
Thus, and in order to reduce the numerical cost, it is a sensible
choice to neglect vertex corrections altogether, and to keep the polarization function on the lowest level, i.e. the random phase
approximation (RPA) which is the convolution product of two non-interacting Green functions in frequency-momentum space. 
The corresponding self-energy is named the $GW^{(0)}$ self-energy and has been introduced by von Barth and Holm \cite{BarthHolm:PRB54_1996}, who
were also the first to study the fully self-consistent $GW$ approximation \cite{Holm:prb57_98}. Throughout this work, the $GW^{(0)}$ self-energy
will be analyzed.

Having been used in solid state physics traditionally, 
the $GW^{(0)}$-method was also applied to study correlations in hot and dense plasmas, recently.
The equation of state \cite{FehrKraeft_CPP35_463_1995,Wierling:CPP38_1998}, as well as optical properties of 
electron hole plasmas in highly excited semiconductors \cite{Schepe-Schmielau:PSSb206_1998}, and 
dense hydrogen plasmas \cite{Fortmann:CPP47_2007} were investigated.  

In general, the calculation of such macroscopic observables of many-particle system involves numerical operations that need the spectral function
as an input. 
Since the self-consistent calculation of the self-energy, 
even in $GW^{(0)}$-approximation, is itself already a numerically demanding task, it is
worth looking for an analytic solution of the $GW^{(0)}$ equations, which reproduces the numerical solution at least in a certain range 
of plasma parameters. Such an analytic expression then also allows to study the self-energy in various limiting cases, such as the low density
limit or the limit of high momenta, which are difficult to access in the numerical treatment. Furthermore, an analytic expression, being already
close to the numerical solution permits the calculation of the full $GW^{(0)}$ self-energy using only few iterations.

Analytical expressions for the single-particle self-energy have already been given by other authors, 
e.g. Fennel and Wilfer \cite{FennelWilfer_AnnPhysL32_265_1974} and Kraeft et al. \cite{KraeftKrempEbelingRoepke1986}. They calculated the
self-energy in first order of the perturbation expansion with respect to the dynamically
screened potential. Besides being far from the converged $GW^{(0)}$ self-energy, their result is independent of density,
i.e. the single-particle life-time is
finite even in vacuum. 
As shown in
\cite{Fortmann_JPhysA41_445501_2008}, this unphysical behaviour is a direct consequence of the perturbative treatment. 
By using a non-perturbative ansatz, an expression for the
self-consistent self-energy in a classical one-component plasma
was presented, that reproduces the full $GW^{(0)}$ self-energy at small momenta,
i.e. for slow particles. The behaviour of the quasi-particle damping at larger momenta remained open and will be investigated in the present
work.
Secondly, based on the information gathered about the low and high momentum behaviour, an interpolation formula will be derived, 
that gives the quasi-particle damping at arbitrary momenta. 

The work is organized in the following way: After a brief outline of the
$GW^{(0)}$-approximation in the next section, numerical results will be given in section \ref{sec:numresults}
for the single-particle spectral function for various sets of parameters electron density $n_\mathrm{e}$ and
electron temperature $T$. In section \ref{sec:analyticexpression} the analytic
expression for the quasi-particle damping width is presented and comparison to the numerical
results are given. Section \ref{sec:chempot} deals with the application of the derived formulae to the calculation of the chemical potential as a
function of density and temperature.
An appendix contains the detailed derivation of the analytic self-energy.
As a model system, we focus on the electron one-component plasma, ions are treated as a homogeneously distributed background of positive charges
(jellium model).

%%%%%%%%%%%%%%%%%%%%%%%%%%%%%%%%%%%%%%%%%%%%%%%%%%%%%%%%%%%%
%
\section{Spectral function and self-energy\label{sec:sf-se}}
We start our discussion with 
the integral equation for the imaginary part of the
single-particle self-energy in $GW^{(0)}$ approximation:
\begin{equation}
  \mathrm{Im}\,\Sigma(\mathbf{p},\omega+\mathrm{i}\delta)= 
  \frac{1}{n_\mathrm{F}(\omega)}\sum_{\mathbf{q}}\int_{-\infty}^{\infty}\frac{\mathrm{d}\omega'}{2\pi}V(q)
  A(\mathbf{p}-\mathbf{q},\omega-\omega')
  \mathrm{Im}\,\epsilon^{-1}_\mathrm{RPA}(\mathbf{q},\omega')\,
  n_\mathrm{B}(\omega')\,n_\mathrm{F}(\omega-\omega')~.
  \label{eqn:sigmacorr_010}
\end{equation}
$V(q)=8\pi/q^2\Omega_0$ is the Fourier transform of the Coulomb potential with the normalization volume $\Omega_0$. It is multiplied by
the inverse dielectric function in RPA,
\begin{equation}
  \epsilon_\mathrm{RPA}(q,\omega)
  =
  1-V(q)\sum_{\mathbf{p}}
  \frac{
  n_\mathrm{F}(\varepsilon_{\mathbf{p}-\mathbf{q}/2})-
  n_\mathrm{F}(\varepsilon_{\mathbf{p}+\mathbf{q}/2})
  }
  {
  \omega+
  \varepsilon_{\mathbf{p}-\mathbf{q}/2}-
  \varepsilon_{\mathbf{p}+\mathbf{q}/2}
  }~.
  \label{eqn:epsPi}
\end{equation}
to account for dynamical screening of the interaction.
Furthermore, the Fermi-Dirac and the Bose-Einstein distribution function,
  $n_\mathrm{F/B}(\varepsilon)=\left[ \exp
  (\varepsilon/k_\mathrm{B}T)\pm1 \right]^{-1}$\,,
were introduced.
Note, that the dielectric function is only determined once, at the beginning of the calculation. In particular, the single-particle energies
$\varepsilon_\mathbf{p}=p^2-\mu_\mathrm{e}$ entering equation (\ref{eqn:epsPi}) are determined from the non-interacting chemical potential, 
whereas during the course of the self-consistent
calculation of the self-energy, the chemical potential is recalculated at each step via inversion of the 
density relation
\begin{equation}
  n_\mathrm{e}(\mu_\mathrm{e},T)=2\sum_{\mathbf{p}}\int\frac{d\omega}{2\pi}A(\mathbf{p},\omega)\,n_\mathrm{F}(\omega)~.
  \label{eqn:densityrelation}
\end{equation}
Using the self-consistent chemical potential also in the RPA polarization function leads to violation of the $f$-sum rule,
i.e. conservation of the number of particles.

The real part of the self-energy is obtained via the Kramers-Kronig relation \cite{Mahan:Book}
%\begin{equation}
%  \mathrm{Re}\,\Sigma(\mathbf{p},\omega)=\Sigma^\mathrm{HF}(\mathbf{p})+\mathcal{P}\int_{-\infty}^{\infty}\frac{d\omega}{\pi}\frac{\mathrm{Im}\,\Sigma(\mathbf{p},\omega'+i\delta)}{\omega-\omega'}~,
%  \label{eqn:KramersKronig_Sigma}
%\end{equation}
%where $\mathcal{P}$ denotes the Cauchy principal value of the integral. 
%$\Sigma^\mathrm{HF}(\mathbf{p})$ is the frequency independent Hartree-Fock part of the
%self-energy,
%\begin{equation}
%  \Sigma^\mathrm{HF}(\mathbf{p})=-\sum_{\mathbf{q}}V(q)\int_{-\infty}^{\infty}\frac{d\omega}{2\pi}A(\mathbf{p}-\mathbf{q},\omega)n_\mathrm{F}(\omega)~.
%  \label{eqn:SigmaHFInteracting}
%\end{equation}
%
%Note, that the spectral function, that depends on the self-energy via Dyson's equation (\ref{eqn:Dyson}) enters both the dynamical part of the
%self-energy (\ref{eqn:sigmacorr_010}) as well as the Hartree-Fock part (\ref{eqn:SigmaHFInteracting}). 
All quantities (spectral function, self-energy, and chemical potential) have to be determined in a self-consistent way. 
This is usually achieved by solving equations (\ref{eqn:Dyson}), (\ref{eqn:sigmacorr_010}), 
%(\ref{eqn:KramersKronig_Sigma}), (\ref{eqn:SigmaHFInteracting}), 
and (\ref{eqn:densityrelation}) iteratively.The numerical algorithm is discussed in detail in Ref.~\cite{Fortmann_JPhysA41_445501_2008}.
%One starts from a suitable ``guess'' for the spectral function or the self-energy and 
%computes successively the imaginary part of the self-energy via equation (\ref{eqn:sigmacorr_010}), the real part via the Kramers-Kronig relation,
%and the chemical potential by inversion of (\ref{eqn:densityrelation}). At the
%end of this cycle, one obtains a new instance of the spectral function via Dyson's equation (\ref{eqn:Dyson}). 
%The cycle is repeated, until the spectral functions of two succeeding
%iterations coincide. Consistency of at least $99.9\%$ is usually achieved after 10 iterations.
%
%To control the numerical algorithm, the results
%are checked against sum-rules, e.g. the normalization integral 
%\begin{equation}
%  \int_{-\infty}^{\infty}\frac{d\omega}{2\pi}\,A(\mathbf{p},\omega)=1~.
%  \label{eqn:Moment_A0}
%\end{equation}
%Other exact relations are known also for the first and second moment of the spectral function, as well as for the normalization of the imaginary
%part of the self-energy, which we refrain from giving here. Details can be found in
%Ref.~\cite{BarthHolm:PRB54_1996}.

%%%%%%%%%%%%%%%%%%%%%%%%%%%%%%%%%%%%%%%%%%%%%%%%%%%%%%%%%%%%%%%%%%%%%%%%%%%%
\section{Numerical results\label{sec:numresults}}
The spectral function was calculated for the case of a hot one-component electron plasma. Temperature and density were chosen such, that the 
plasma degeneracy parameter
\begin{equation}
  \theta=\frac{T}{E_\mathrm{F}}~,
  \label{eqn:ThetaDegeneracy}
\end{equation}
is always larger than 1, i.e. the plasma is non-degenerate. 
Furthermore, the temperature is fixed above
the ionization energy of hydrogen, $T\gg 1 \,\mathrm{Ry}$, such that bound states can be neglected in the calculations. At lower temperature,
bound states have to be include, e.g. via the t-matrix. For an application in electron-hole plasmas, see reference
\cite{Schepe-Schmielau:PSSb206_1998}. 
The electron density is adjusted such that the
the plasma coupling parameter
\begin{equation}
  \Gamma=\frac{2}{T}\left(\frac{4\pi n_\mathrm{e}}{3} \right)^{1/3}~,
  \label{eqn:Gamma}
\end{equation}
which gives the mean Coulomb interaction energy compared to the thermal energy, is smaller than 1 in all calculations, i.e. we are in
the limit of weak coupling. 

\begin{figure}[ht]
  \begin{center}
    \includegraphics[width=.5\textwidth,angle=0,clip,draft=0]{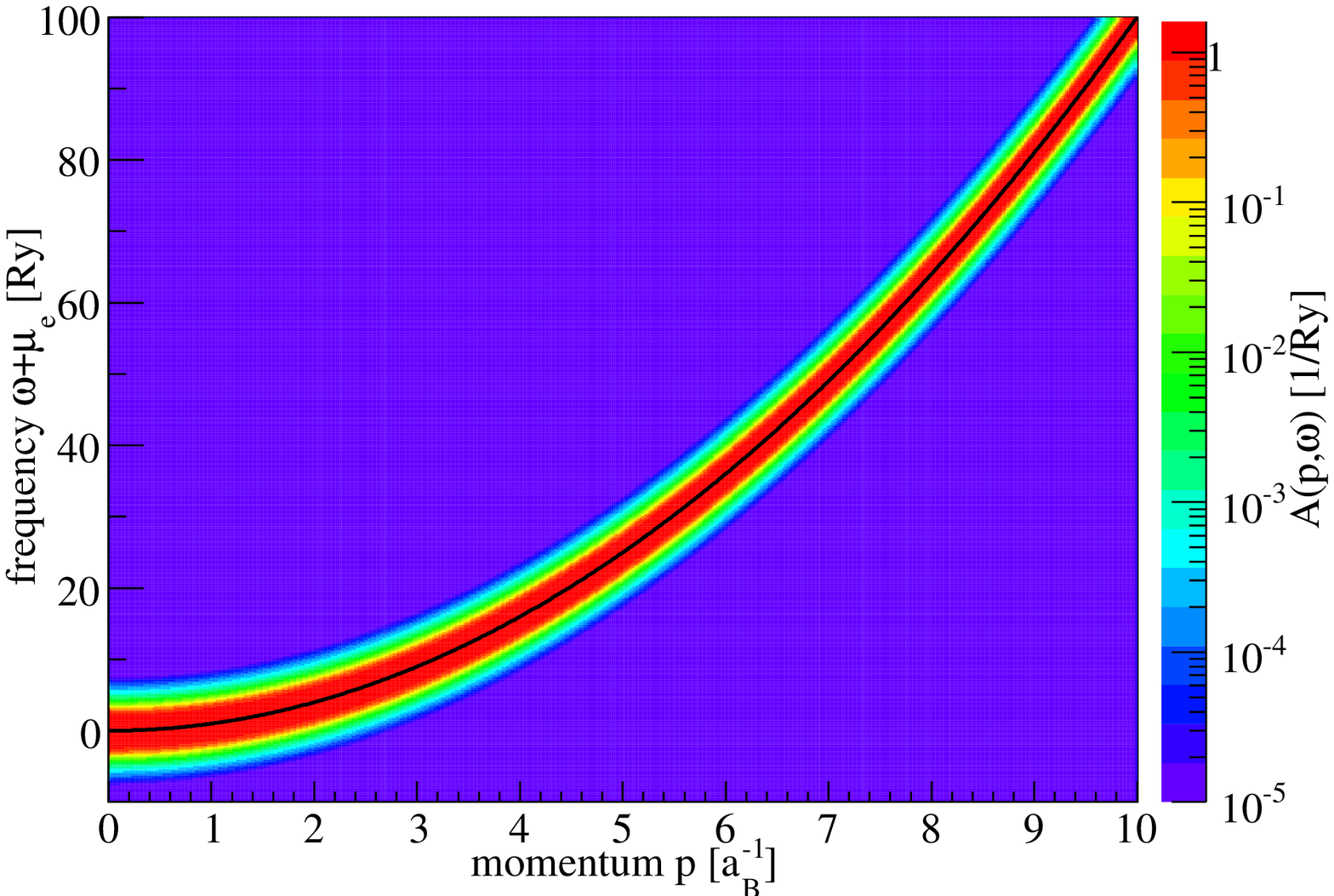}
    \includegraphics[width=.5\textwidth,angle=0,clip]{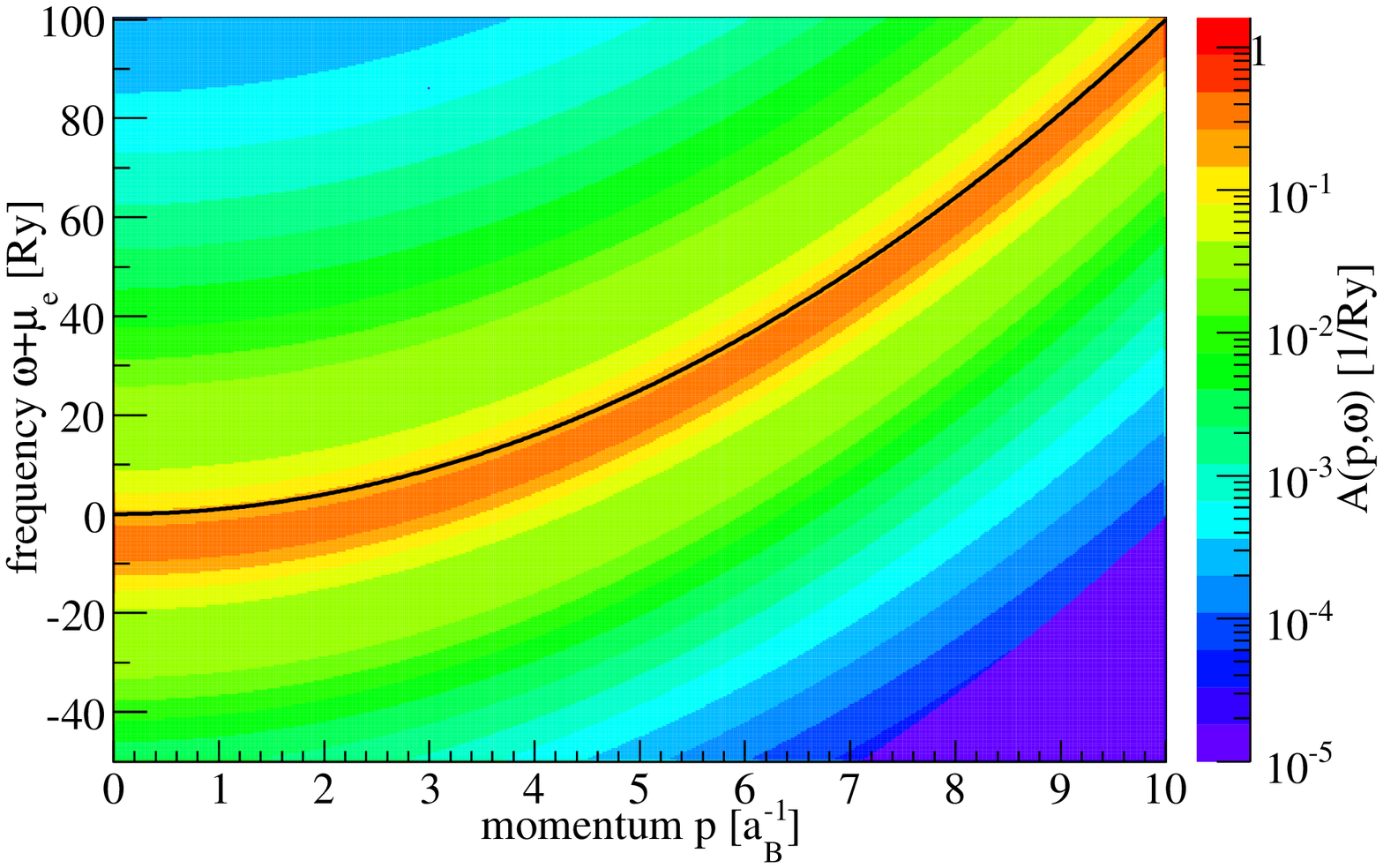}
  \end{center}
  \caption{(Color online) Contour plots of the spectral function
  as a function of momentum and frequency. The colour scale is
  logarithmic.
  Results are shown for two different densities ($n=7\times
  10^{21}\,\mathrm{cm^{-3}}$, upper graph) and ($n=7\times
  10^{25}\,\mathrm{cm^{-3}}$, lower graph). For these parameters, 
  the plasma coupling parameter is
  $\Gamma=4.4\times 10^{-3}$, and $\Gamma=9.6\times 10^{-2}$,
  respectively. The degeneracy parameter is $\theta=7.5\times
  10^{2}$ and $\theta=1.6$, respectively. The black line
  indicates the free particle dispersion $\omega=\varepsilon_p$.
  \label{fig:sf_3d_n7e+21}
  }
\end{figure}

In figure \ref{fig:sf_3d_n7e+21}, we show contour plots
of the spectral function in frequency and
momentum space for two different densities ($n_\mathrm{e}=7\times
10^{21}\,\mathrm{cm^{-3}}$, upper graph) and ($n_\mathrm{e}=7\times
10^{25}\,\mathrm{cm^{-3}}$, lower graph). The temperature is set
to $T=1000\,\mathrm{eV}$ in both calculations.
The free particle dispersion $\omega=\varepsilon_p$ is shown as solid black line.

In the first case, the plasma is classical ($\theta=7.5\times 10^{2}$) and weakly coupled ($\Gamma=4.4\times 10^{-3}$). The spectral function
is symmetrically broadened and the maximum is found at the free dispersion, i.e. there is no notable quasi-particle
shift at the present conditions.  At increasing momentum, the width of the spectral function decreases, and the maximum value increases; the
norm is preserved.

The situation changes, when we go to higher densities, cf. the lower graph in figure \ref{fig:sf_3d_n7e+21}. The chosen parameters are typical
solar core parameters \cite{Bahcall:RMP67.781.1995}. The degeneracy parameter is now $\theta=1.6$ and the coupling parameter is $\Gamma=0.096$.
The increased degeneracy and coupling result in a significant modification of the spectral function as compared to the low density case:
A shift of the spectral function's maximum to smaller frequencies is
observed, the Hartree-Fock shift. The shift due to dynamic correlations is still small at the present
conditions; it becomes important in strongly degenerate systems \cite{BarthHolm:PRB54_1996}. 
Furthermore, shoulders appearing in the wings of the
main quasi-particle peak at small momenta, indicate the excitation of new quasi-particles, so-called plasmarons
\cite{Lundquist_PhysKondMat6_193_1967}. They can be seen more clearly in figure~\ref{fig:sf_GW-Gauss_n7e25_T1000} (solid curve). The plasmaron
satellites are separated from the main peak by roughly the plasma frequency
$\omega_\mathrm{pl}=4\sqrt{\pi n_\mathrm{e}}$ which is about $23\,\mathrm{Ry}$ at the present density. In the former case of lower density
no plasmarons appear, only a featureless, single resonance is obtained.
At higher momenta, the plasmarons merge into the central peak. As in the low-density case, the position of the
maximum approaches the single-particle dispersion, due to the decreasing Hartree-Fock shift at high $p$. 
Again, the width of the spectral
function decreases with increasing momentum. 
%\begin{figure}[ht]
%  \begin{center}
%    \includegraphics[width=.5\textwidth,angle=0,clip]{sfcuts_n7e+25_T1000eV}
%    \includegraphics[width=.5\textwidth,angle=0,clip]{sfcontour_n7e+25_T1000eV}
%  \end{center}
%  \caption{(Color online) Three dimensional representations of the spectral function for solar core plasma parameters ($n=7\times 10^{25}\,\mathrm{cm^{-3}}$,
%  $T=1000\,\mathrm{eV}$).  The upper graph shows cuts through the spectral function at various momenta
%  $p$, the lower graph shows a contour plot of the spectral function. Here, the z-axis is scaled logarithmically. For the present parameters, the plasma coupling parameter is
%  $\Gamma=9.6\times 10^{-2}$, the degeneracy parameter is $\theta=1.6$.}
%  \label{fig:sf_3d_n7e+25}
%\end{figure}

This is visible more clearly in figures \ref{fig:sf_GW-Gauss_n7e19} - \ref{fig:sf_GW-Gauss_n7e25_T1000}. Here,
the solid curves represent the $GW^{(0)}$
spectral function as a function of frequency. Results are shown for three different momenta, $p=0\,a_\mathrm{B}^{-1}$ (a),
$p=50\,a_\mathrm{B}^{-1}$ (b), and
$p=100\,a_\mathrm{B}^{-1}$ (c). Two different temperatures are considered, $T=100\,\mathrm{eV}$ (figures
\ref{fig:sf_GW-Gauss_n7e19}-\ref{fig:sf_GW-Gauss_n7e21}) and $T=1000\,\mathrm{eV}$ (figures
\ref{fig:sf_GW-Gauss_n7e21_T1000}-\ref{fig:sf_GW-Gauss_n7e25_T1000}) and for each temperature three different densities are studied.
With increasing momentum $p$,
the spectral function becomes more and more narrow,
converging eventually into a narrow on-shell resonance, located at the unperturbed single-particle dispersion $\omega+\mu_\mathrm{e}=p^2$.

As a general feature, one can observe an increase of the spectral function's width with increasing density and
with increasing temperature. The increase with density is due to the increased coupling, while the increase with temperature reflects the
thermal broadening of the momentum distribution function $n_\mathrm{F}(\omega)$ that enters the self-energy and thereby also the spectral
function. From these results, we see that the spectral function has a quite simple form in the limit of low coupling, i.e. at low densities and high
temperatures.

The numerical results are compared to a Gaussian ansatz for the spectral function, shown as the dashed curve in figures
\ref{fig:sf_GW-Gauss_n7e19} - \ref{fig:sf_GW-Gauss_n7e25_T1000}. The explicit form of the Gaussian is given as equation (\ref{eqn:GaussDef}),
below. It's sole free parameter is the width, denoted by $\sigma_p$. An
analytic expression for $\sigma_p$ will be derived in section~\ref{sec:analyticexpression}. The coincidence is in general good at high momenta,
whereas at low momenta, the spectral function deviates from the Gaussian. In particular, the steep wings and the smoother plateau that form at low
momenta are not reproduced by the Gaussian. Also, the plasmaron peaks appearing in the spectral function at high density (see figure
\ref{fig:sf_GW-Gauss_n7e25_T1000}, cannot be described by the single Gaussian.

Determining $\sigma_p$ via least-squares fitting of the Gaussian ansatz to the numerical data at each $p$ leads to the solid curve in 
figure \ref{fig:sigma_p_pade-gwfit}), obtained in the case of $n=7\times 10^{20}\,\mathrm{cm}^{-3}$ and $T=100\,\mathrm{eV}$. Starting at some
finite value $\sigma_0$ at $p=0$, the width falls off slowly towards larger $p$. The dashed curve shows $\sigma_p$ as obtained from the analytic
formula that will be derived in the now following section.

%%%%%%%%%%%%%%%%%%%%%%%%%%%%%%%%%%%%%%%%%%%%%%%%%%%%%%%%%%%%%%%%%%%%%%%%%%%%%%%%%%%
%
\section{Analytical expression for the quasi-particle self-energy\label{sec:analyticexpression}}
The solution of the $GW^{(0)}$-equation (\ref{eqn:sigmacorr_010}) requires a considerable numerical effort. So far (see e.g. the work by 
Fennel and Wilfer in \cite{FennelWilfer_AnnPhysL32_265_1974}), 
attempts to solve the integral
(\ref{eqn:sigmacorr_010}) analytically were led by the idea to replace the spectral function
on the r.h.s. by its non-interacting counterpart, $A^\mathrm{(0)}(\mathbf{p},\omega)=2\pi\delta(\varepsilon_\mathbf{p}-\omega)$, i.e. to go back to the
perturbation expansion of the self-energy and neglect the implied self-consistency. At the same time, the inverse dielectric function is usually 
replaced by a simplified expression, e.g. the Born approximation or the plasmon-pole approximation \cite{KraeftKrempEbelingRoepke1986}. 
Whereas the second simplification is indispensable due to the
complicated structure of the inverse dielectric function, the first one, i.e. the recursion to
the quasi-particle picture, is not necessary,
as was shown by the author in Ref.~\cite{Fortmann_JPhysA41_445501_2008}. In fact, the result that one obtains in the quasi-particle approximation
is far from the converged result, at least in the high temperature case. Secondly, using the quasi-particle approximation, the imaginary part of  
the self-energy is not density dependent,
i.e. a finite life-time of the particle states is obtained even in vacuum.
This unphysical result can only be overcome if one sticks to the self-consistency of the self-energy, i.e. if one leaves the imaginary part of
the self-energy entering the r.h.s. of equation (\ref{eqn:sigmacorr_010}) finite.

Using the statically screened Born approximation, which describes the binary collision among electrons via a statically screened potential,
a scaling law $\mathrm{Im}\,\Sigma(\mathbf{p},\omega^\mathrm{QP}(\mathbf{p}))\propto \Gamma^{3/4}$ was found \cite{Fortmann_JPhysA41_445501_2008}. Hence, the spectral functions
width vanishes when the plasma coupling
parameter $\Gamma$ (see equation (\ref{eqn:Gamma})) tends to 0. 
An expression for the self-energy was found, that reproduces the converged $GW^{(0)}$ calculations at small
momenta, $p\ll \kappa$. At higher momenta, the derived expression ceases to be valid.

In this work, a different approximation to the dielectric function is studied, namely the plasmon-pole approximation
\cite{KraeftKrempEbelingRoepke1986}. This means, that the inverse
dielectric function is replaced by a sum of two delta-functions that describe the location of the plasmon resonances,
\begin{equation}
  \mathrm{Im}\,\epsilon_\mathrm{RPA}^{-1}(\mathbf{q},\omega')\longrightarrow \mathrm{Im}\,\epsilon_\mathrm{PPA}^{-1}(\mathbf{q},\omega')=
  -\frac{\pi}{2}\frac{\omega_\mathrm{pl}^2}{\omega_\mathbf{q}}\left[ \delta(\omega-\omega_\mathbf{q})+\delta(\omega+\omega_\mathbf{q}) \right]~.
  \label{eqn:plasmonpole}
\end{equation}
For classical plasmas, the plasmon dispersion $\omega_\mathbf{q}$ can be approximated by the Bohm-Gross dispersion relation
\cite{GrossBohm_PhysRev.75.1851},
\begin{equation}
  \omega_\mathbf{q}^2=\omega_\mathrm{pl}^2\left( 1+\frac{q^2}{\kappa^2} \right)+q^4~.
  \label{eqn:GrossBohm}
\end{equation}
Many-particle and quantum effects on the plasmon dispersion have recently been studied in
\cite{Thiele_PRE_accepted}.

The plasmon-pole approximation allows to perform the frequency integration in equation (\ref{eqn:sigmacorr_010}), resulting in the expression
\begin{multline}
  \mathrm{Im}\,\Sigma(\mathbf{p},\omega+i\delta)= 
\frac{\omega_\mathrm{pl}^2}{4}\sum_{\mathbf{q}}V(q)\frac{1}{\omega_\mathbf{q}}\bigg[
  A(\mathbf{p}-\mathbf{q},\omega-\omega_\mathbf{q})
  n_\mathrm{B}(\omega_\mathbf{q})\exp(\omega_\mathbf{q}/T)-
  A(\mathbf{p}-\mathbf{q},\omega+\omega_\mathbf{q})
  n_\mathrm{B}(-\omega_\mathbf{q})\exp(-\omega_\mathbf{q}/T)\bigg]
  ~.
  \label{eqn:ImSigmaPPA}
\end{multline}
We will first study the case of high momenta, i.e. momenta that are large against any other momentum scale or inverse length
scale, such as the mean momentum with respect to the Boltzmann distribution, $\bar p=\sqrt{3T/2}$
or the inverse screening length $\kappa=\sqrt{8\pi n/T}$.

As discussed in the previous section,
the numerical results for the spectral function
at high momenta can well be reproduced by a Gaussian. Thus, we make the following ansatz for the spectral function:
\begin{equation}
  A_\mathrm{Gauss}(\mathbf{p},\omega)=-\frac{\sqrt{2\pi}}{\sigma_p}\exp\left(
  -\frac{(\omega-\varepsilon_\mathbf{p}-\Sigma^\mathrm{HF}(\mathbf{p}))^2}{2\sigma_p^2} \right)~.
  \label{eqn:GaussDef}
\end{equation}
Note, that only the Hartree-Fock contribution to the real part of the self-energy appears, the frequency dependent part is usually small near the
quasi-particle dispersion, 
\begin{equation}
  \omega^\mathrm{QP}(\mathbf{p})=\varepsilon_\mathbf{p}+\mathrm{Re}\,\Sigma(\mathbf{p},\omega)\left.\right|_{\omega=\omega^\mathrm{QP}(\mathbf{p})}~,
\end{equation}
which therefore can be approximated as $\omega^\mathrm{QP}(\mathbf{p})=\varepsilon_\mathbf{p}+\mathrm{Re}\,\Sigma^\mathrm{HF}(\mathbf{p})$.
In the following, we make use of the knowledge about the width parameter $\sigma_p$ that we gathered already through simple least-squares fitting of the Gaussian ansatz to
the spectral function in order
to solve the integrals in equation (\ref{eqn:ImSigmaPPA}).

\begin{figure}[ht]
  \begin{center}
    \includegraphics[width=.5\textwidth,angle=0,clip]{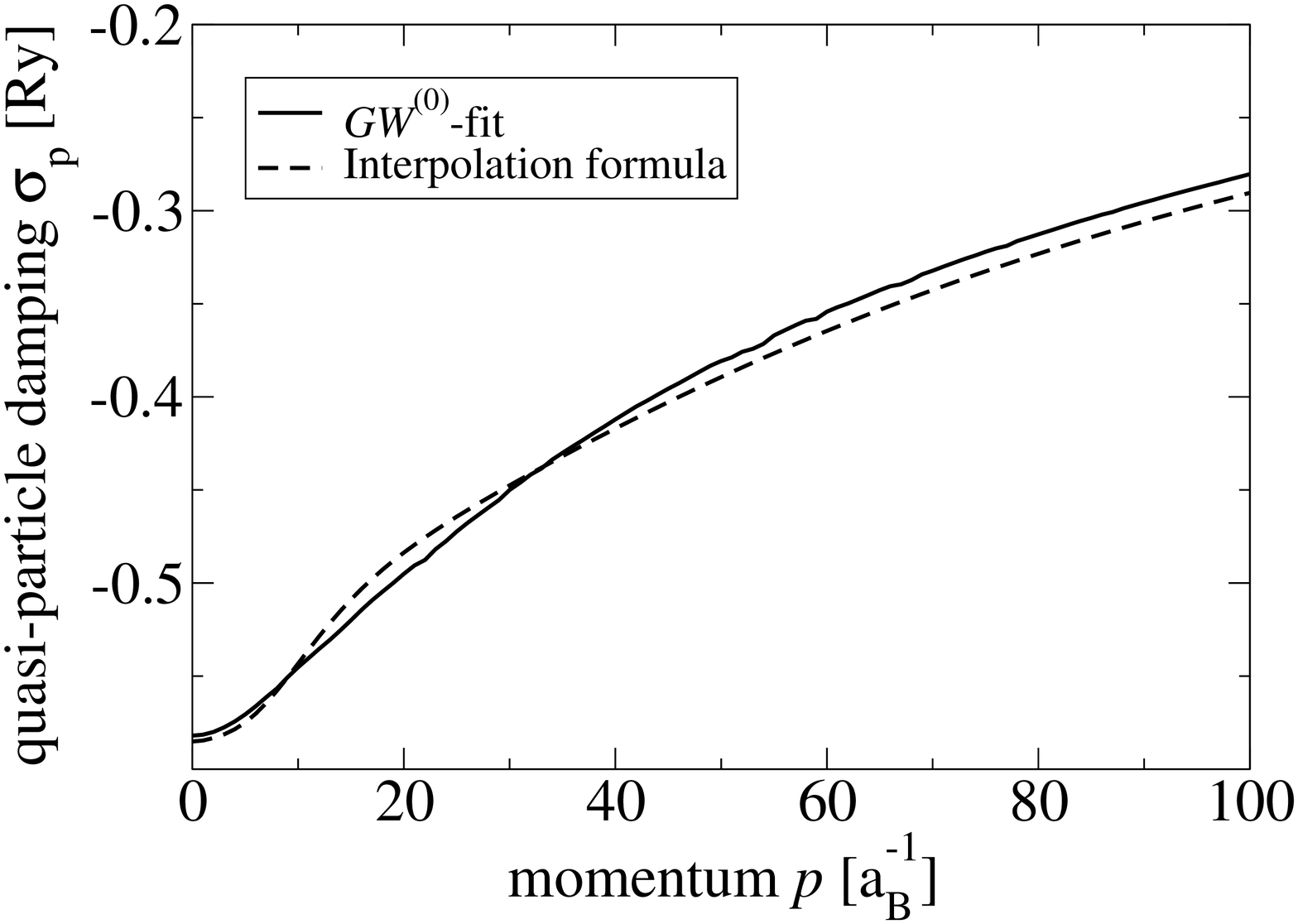}
  \end{center}
  \caption{Effective quasi-particle damping $\sigma_p$ as a function of momentum $p$ for plasma density $n=7\times 10^{20}\,\mathrm{cm^{-3}}$ and
  temperature $T=100\,\mathrm{eV}$. The fit-parameters for the Gaussian fit to the full $GW^{(0)}$-calculations are given as dots, the solid line
  denotes the analytic interpolation formula (\ref{eqn:sigma_p_interpolation}).}
  \label{fig:sigma_p_pade-gwfit}
\end{figure}

First, we replace the spectral function on
the r.h.s by the Gaussian ansatz (\ref{eqn:GaussDef}) and evaluate the emerging equation at the single-particle dispersion
$\omega^\mathrm{QP}(\mathbf{p})$.
By claiming that the Gaussian and the spectral function have the same value at the
quasi-particle energy, we identify
$\sigma_p=\sqrt{\pi/2}\mathrm{Im}\,\Sigma(\mathbf{p},\omega^\mathrm{QP}(\mathbf{p})))$.
Figure \ref{fig:sigma_p_pade-gwfit} shows, that the quasi-particle damping $\sigma_p$ is a smooth function of $p$, that varies only little on the scale of the screening
parameter $\kappa$. Since the latter defines the scale on which contributions to the $q$-integral are most important, we can neglect the momentum
shift in the self-energy on the r.h.s., i.e. we can replace the spectral function on the r.h.s of equation
(\ref{eqn:ImSigmaPPA}) as
\begin{equation}
  A(\mathbf{p}-\mathbf{q},\varepsilon_\mathbf{p}+\Sigma^\mathrm{HF}(\mathbf{p})\pm\omega_\mathbf{q})\longrightarrow -\frac{\sqrt{2\pi}}{\sigma_p}\exp\left(
  -\frac{(\varepsilon_\mathbf{p}\pm\omega_\mathbf{q}-\varepsilon_{\mathbf{p}-\mathbf{q}})^2}{2\sigma_p^2} \right)~,
  \label{eqn:A-Gauss}
\end{equation}
and can now perform the integral over the angle $\theta$ between the momenta $\mathbf{p}$ and $\mathbf{q}$,
\begin{multline}
  \frac{\sqrt{2\pi}}{\sigma_p}\int_{-1}^{1}d\cos\theta \exp\left( 
  -\frac{(\varepsilon_{\mathbf{p}}\pm\omega_\mathbf{q}-p^2-q^2+2pq\cos\theta+\mu_\mathrm{e})^2}{2\sigma_p^2}
  \right)
  =\\
  \frac {\pi}{2pq}\left (\text {Erf}\left[\frac {(p + q)^2 - p^2\mp\omega_\mathbf{q}} {\sqrt {2} \sigma_p} \right] -
  \text {Erf}\left[\frac {(p - q)^2 -  p^2\pm\omega_\mathbf{q}} {\sqrt {2} \sigma_p} 
  \right] \right)~.
  \label{eqn:thetaIntegration}
\end{multline}

The remaining integration over the modulus of the transfer momentum $q$ can be performed after some further approximations, explained in detail
in appendix \ref{app:GaussPPA}. For large $p$, one finally obtains the transcendent equation

\begin{equation}
  \begin{split}
  \sigma_p&=-1.3357\sqrt{\frac{\pi}{2}}\frac{\omega_\mathrm{pl}}{2p}\left[
  n_\mathrm{B}(\omega_\mathrm{pl})\,\exp(\omega_\mathrm{pl}/T)-n_\mathrm{B}(-\omega_\mathrm{pl})\,\exp(-\omega_\mathrm{pl}/T) \right]
  -\sqrt{\frac{\pi}{2}}\frac{T}{2p}\ln(\kappa^2p^2/\sigma_p^2)
  \end{split}
  \label{eqn:deriv0080}
\end{equation}
The solution of this equation can be expanded for large arguments of the logarithm,
yielding
\begin{equation}
  \begin{split}
  \sigma_p&=-\sqrt{\frac{\pi}{2}}\frac{T}{p}\varphi(p)~,\quad
  \varphi(p)=\Bigg[\xi(p)-\ln \xi(p) +\frac{\ln \xi(p)}{\xi(p)}-\frac{\ln \xi(p)}{\xi^2(p)}+\frac{\ln \xi(p)}{\xi(p)}-\frac{3\ln^2
  \xi(p)}{2\xi^3(p)}+\frac{\ln^2\xi(p)}{2\xi^2(p)}+\frac{\ln^3\xi(p)}{3\xi^3(p)}\Bigg]+\mathcal{O}(p)^{-3}
  \\
  \xi(p)&=\ln(\sqrt{\frac{2}{\pi}}\kappa p^2\exp(A/T)/T)~,\quad
  A=-1.3357\frac{\omega_\mathrm{pl}}{2}\left[n_\mathrm{B}(\omega_\mathrm{pl})\exp(\omega_\mathrm{pl}/T)-n_\mathrm{B}(-\omega_\mathrm{pl})\exp(-\omega_\mathrm{pl}/T)\right]~.
  \end{split}
  \label{eqn:finalsolution}
\end{equation}
Equation (\ref{eqn:finalsolution}) is a solution of equation (\ref{eqn:deriv0080}) provided the argument of the inner
logarithm is larger than Euler's constant $e$, i.e. $\sqrt{\frac{2}{\pi}}\kappa p^2\exp(A/T)/T>e$, i.e. at large $p$. The case of small $p$, where
the previous inequality does not hold, has to be treated separately, see appendix~\ref{app:Pade}.

Together with an expression for the quasi-particle damping at vanishing momentum taken from \cite{Fortmann_JPhysA41_445501_2008} and scaled such that
the maximum of the spectral function at $p=0$ is reproduced, $\sigma_0=-\pi\sqrt{\kappa T}/2$,
an interpolation formula (Pad\'e formula) was derived that covers the complete $p$-range:
\begin{equation}
  \begin{split}
  \sigma_p^\text{Pad\'e}&=\frac{a_0+a_1 p}{1+b_1p+b_2p^2}\tilde\varphi(p)~,\quad
  \tilde\varphi(p)=\Bigg[\tilde\xi(p)-\ln \tilde\xi(p) +\frac{\ln \tilde\xi(p)}{\tilde\xi(p)}-\frac{\ln \tilde\xi(p)}{\tilde\xi^2(p)}+\frac{\ln \tilde\xi(p)}{\tilde\xi(p)}-\frac{3\ln^2
  \tilde\xi(p)}{2\tilde\xi^3(p)}+\frac{\ln^2\tilde\xi(p)}{2\tilde\xi^2(p)}+\frac{\ln^3\tilde\xi(p)}{3\tilde\xi^3(p)}\Bigg]\\
  \tilde \xi(p)&=\ln(e+\sqrt{\frac{2}{\pi}}\kappa p^2\exp(A/T)/T)~,\\
  a_0&=-\frac{\pi}{2}\sqrt{\kappa T}\,,	\qquad a_1=-\kappa\left( \frac{\pi}{2} \right)^{3/2}\,,\qquad
  b_1=\sqrt{\frac{\pi\kappa}{2T}}\,,	\qquad b_2=\frac{\pi\kappa}{2T}~.
\end{split}
\label{eqn:sigma_p_interpolation}
\end{equation}
The function $\tilde{\xi}(p)$ in the last equation differs from $\xi(p)$ in equation (\ref{eqn:finalsolution}) in that Euler's constant $e\simeq 2.7183$ has
been added to the argument of the logarithm. In this way, the function $\tilde{\varphi}(p)$ is regularized at small $p$ and tends to 1 at $p=0$, 
i.e. the quasi-particle damping goes to the correct low-$p$ limit. 
At large $p$, this modification is insignificant, since the original argument rises as $p^2$.
For the detailed derivation, see appendix \ref{app:Pade}.

Expression (\ref{eqn:sigma_p_interpolation}), used in the Gaussian ansatz (\ref{eqn:GaussDef}), leads to a spectral function that well reproduces the numerical data from full
$GW^{(0)}$-calculations: 
Figure \ref{fig:sigma_p_pade-gwfit} (dashed curve) shows the effective quasi-particle damping width $\sigma_p$ as a function of momentum
$p$ for the case $n=7\times 10^{20}\,\mathrm{cm^{-3}}$ and $T=100\,\mathrm{eV}$. The solid curve gives the best-fit value for $\sigma_p$ obtained via
least-square fitting of the full $GW^{(0)}$-calculations assuming the Gaussian form (\ref{eqn:GaussDef}), see section \ref{sec:numresults}.
Both curves coincide to a large extent. 
The largest deviations are observed in the range of $p\simeq 20\,a_\mathrm{B}^{-1}$. At this point, the validity of expression
(\ref{eqn:finalsolution}) as the solution of equation (\ref{eqn:deriv0080}) ceases, 
since the argument of the logarithm becomes smaller than
$e$. As already mentioned, we circumvented this problem by regularizing the logarithms, adding $e$ to its argument. 
The deviation at $p\simeq 20\,a_\mathrm{B}^{-1}$ of up to 15\% is a residue of this procedure.
At higher momenta, the deviation
is generally smaller than $10\%$ and the analytic formula evolves parallel to the fit parameters. 

\begin{figure}[ht]
  \begin{center}
    \subfigure[$p=0$]{\includegraphics[width=.3\textwidth,angle=0,clip]{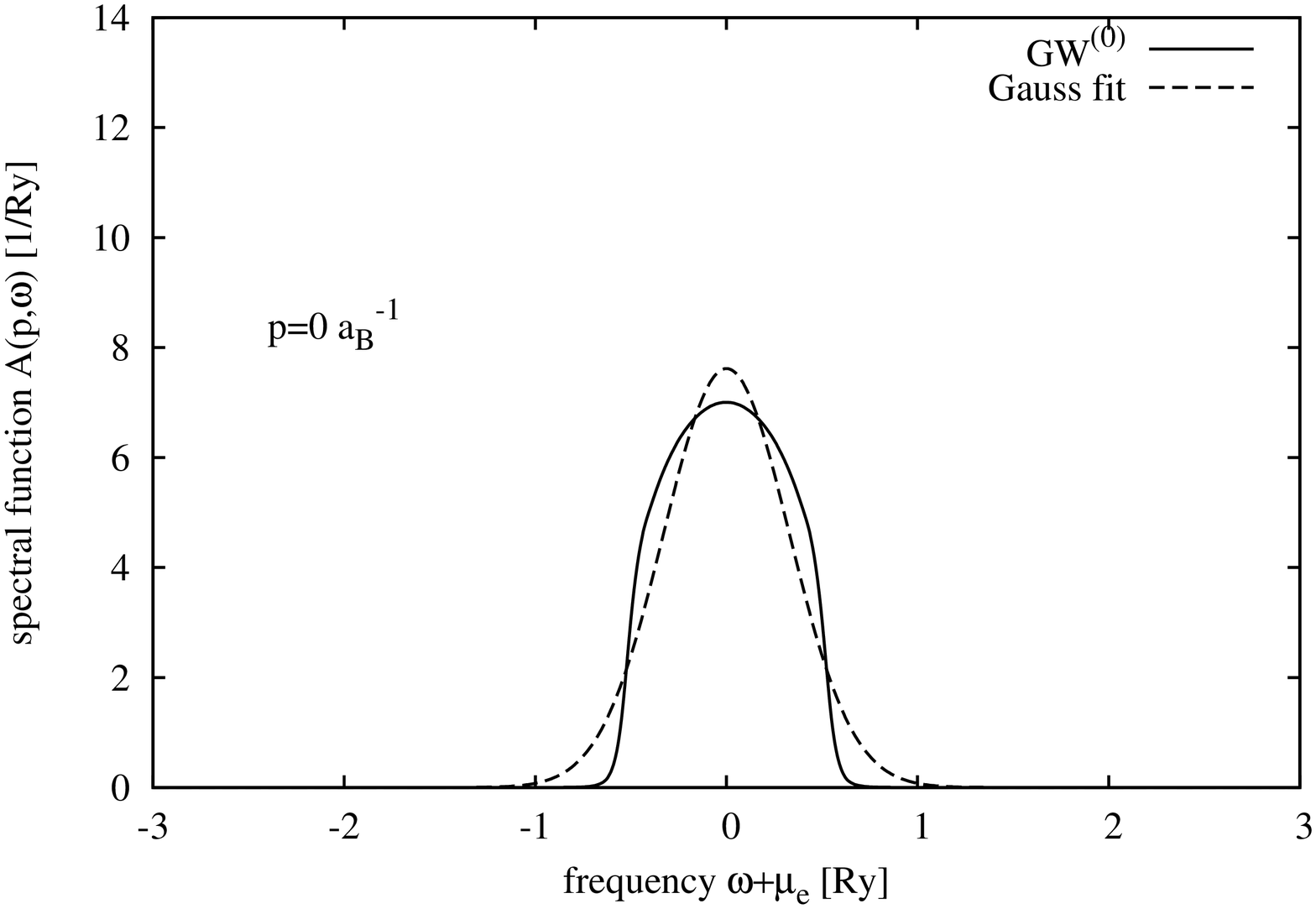}}%
    \subfigure[$p=50\,a_\mathrm{B}^{-1}$]{\includegraphics[width=.3\textwidth,angle=0,clip]{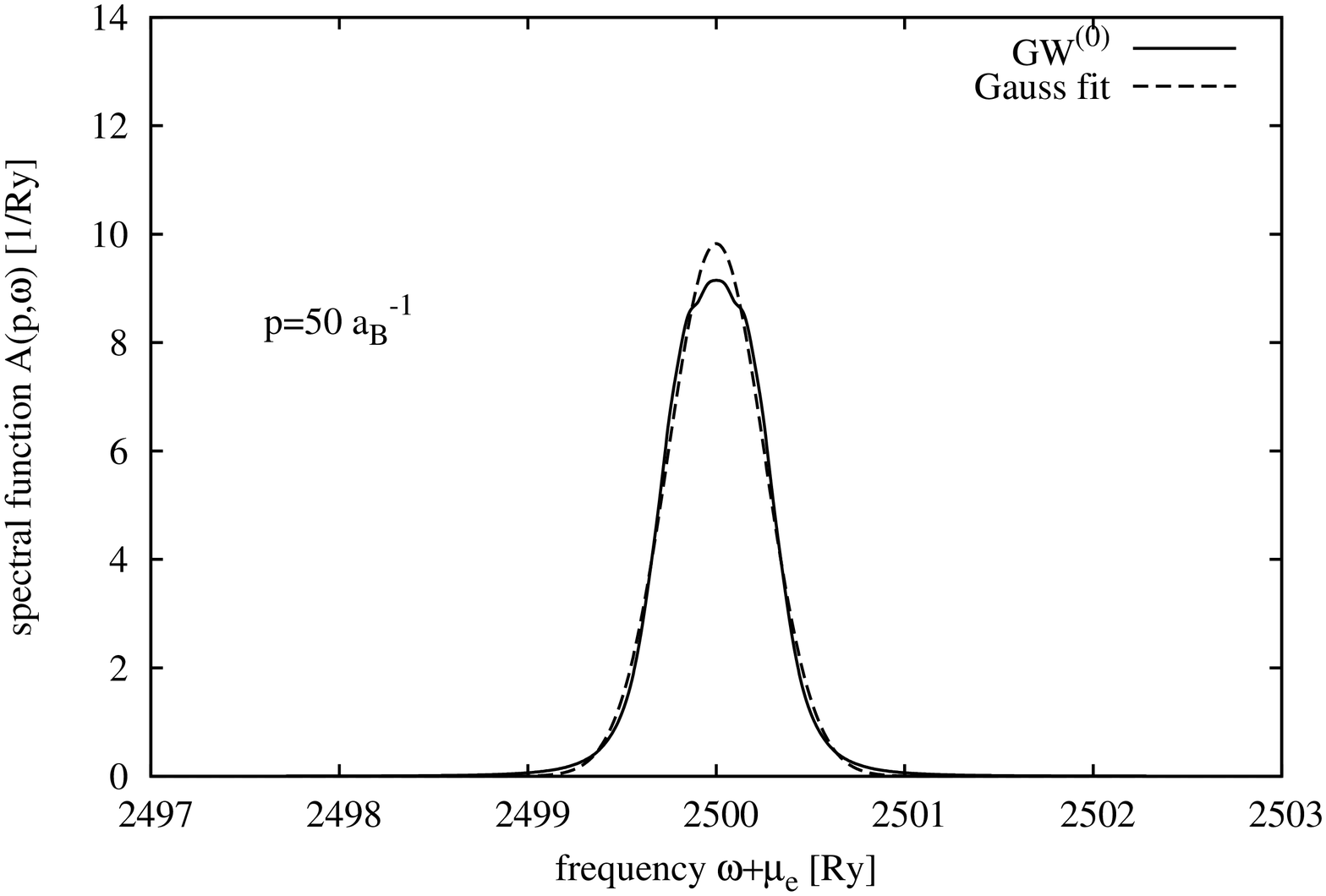}}%
    \subfigure[$p=100\,a_\mathrm{B}^{-1}$]{\includegraphics[width=.3\textwidth,angle=0,clip]{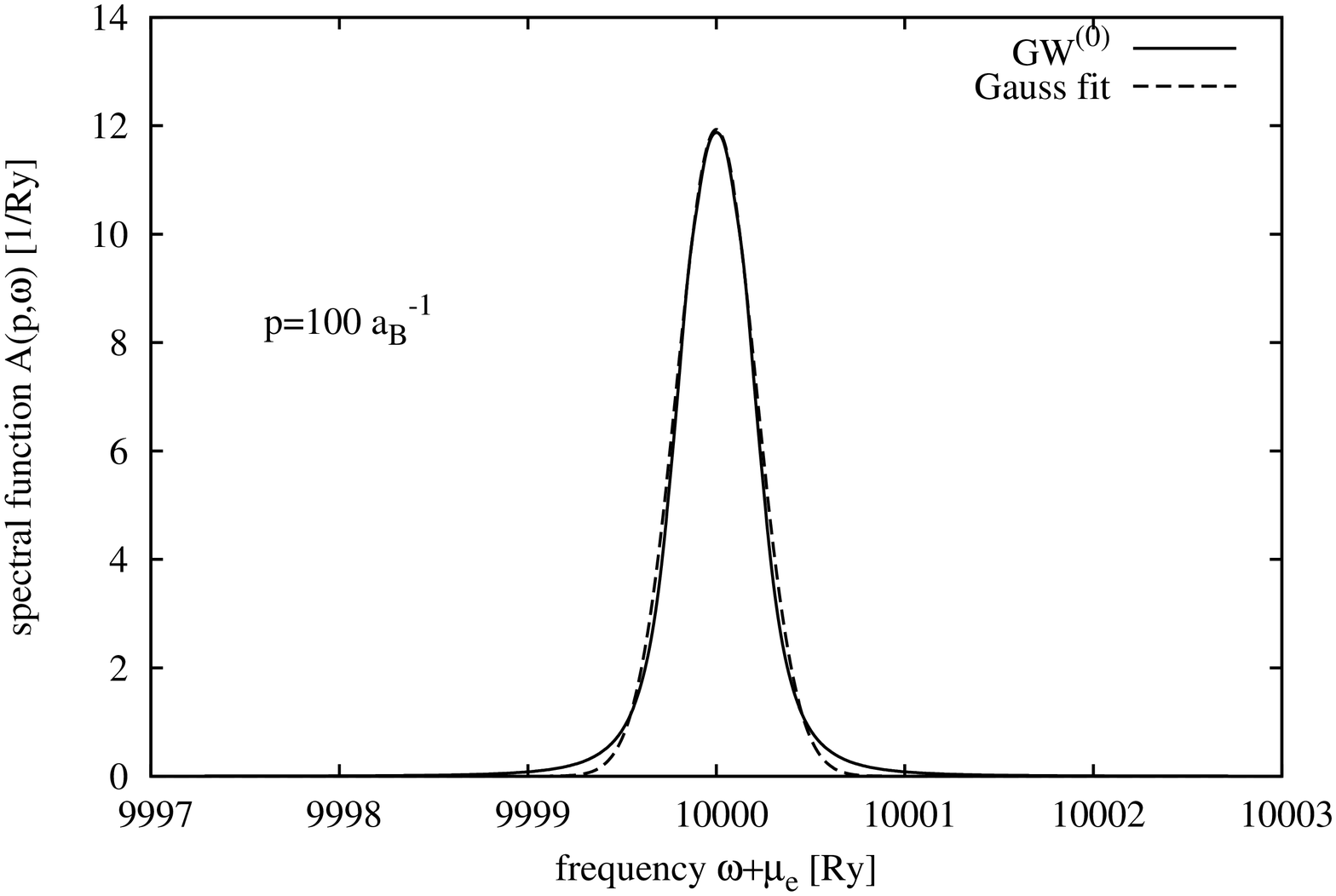}}%
  \end{center}
  \caption{Spectral function in $GW^{(0)}$-approximation (solid lines) and Gaussian ansatz  (dashed lines) with quasi-particle damping width $\sigma_p$ taken from
  equation (\ref{eqn:sigma_p_interpolation}) for three different momenta. Plasma
  parameters: $n=7\times 10^{19}\,\mathrm{cm^{-3}}$, $T=100\,\mathrm{eV}$. The plasma coupling parameter is
  $\Gamma=1.0\times 10^{-2}$, the degeneracy parameter is $\theta=1.6\times 10^{3}$, the Debye screening parameter is $\kappa=6.0\times
  10^{-3}\,a_\mathrm{B}^{-1}$.}
  \label{fig:sf_GW-Gauss_n7e19}
\end{figure}
\begin{figure}[ht]
  \begin{center}
    \subfigure[$p=0$]{\includegraphics[width=.3\textwidth,angle=0,clip]{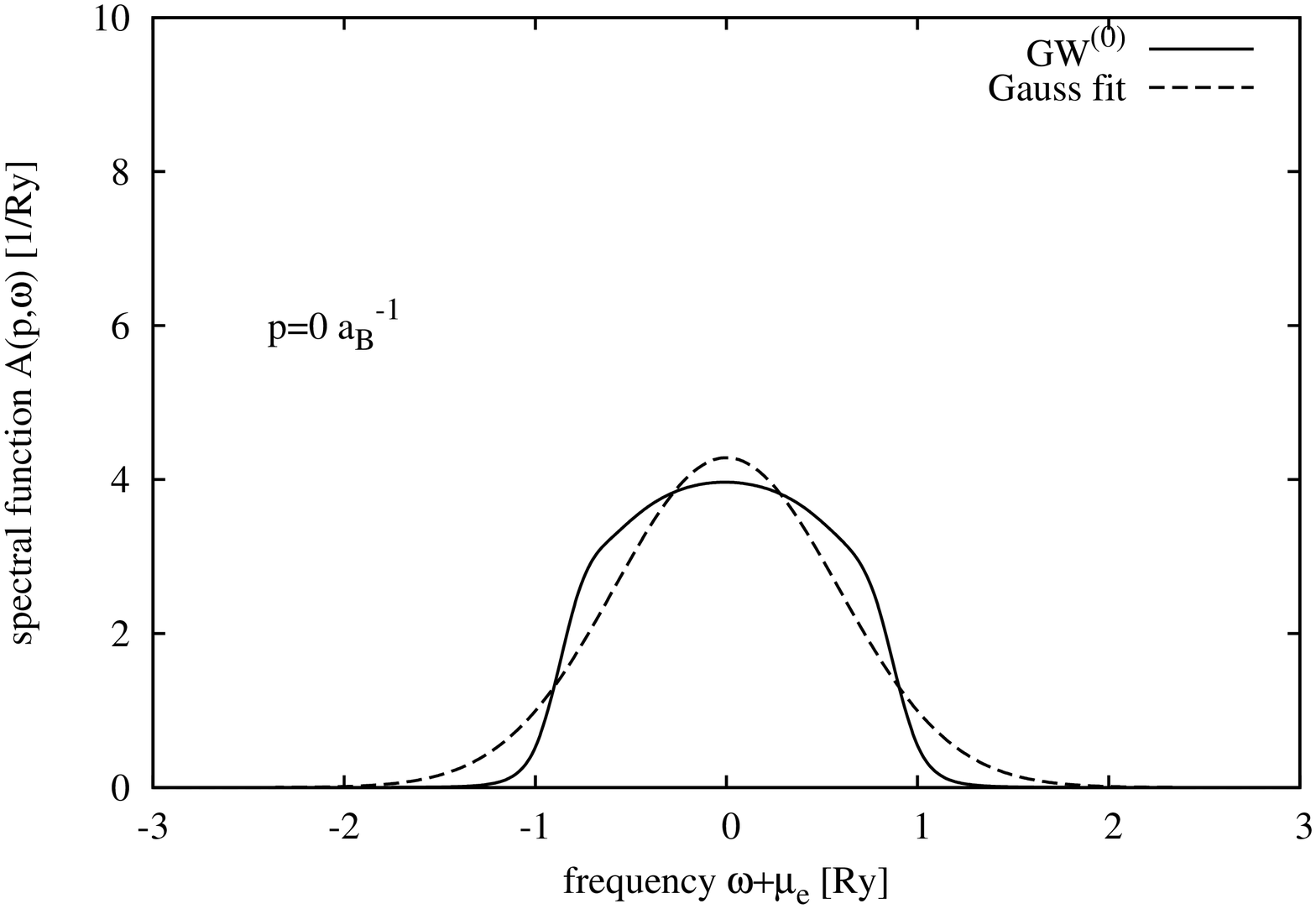}}%
    \subfigure[$p=50\,a_\mathrm{B}^{-1}$]{\includegraphics[width=.3\textwidth,angle=0,clip]{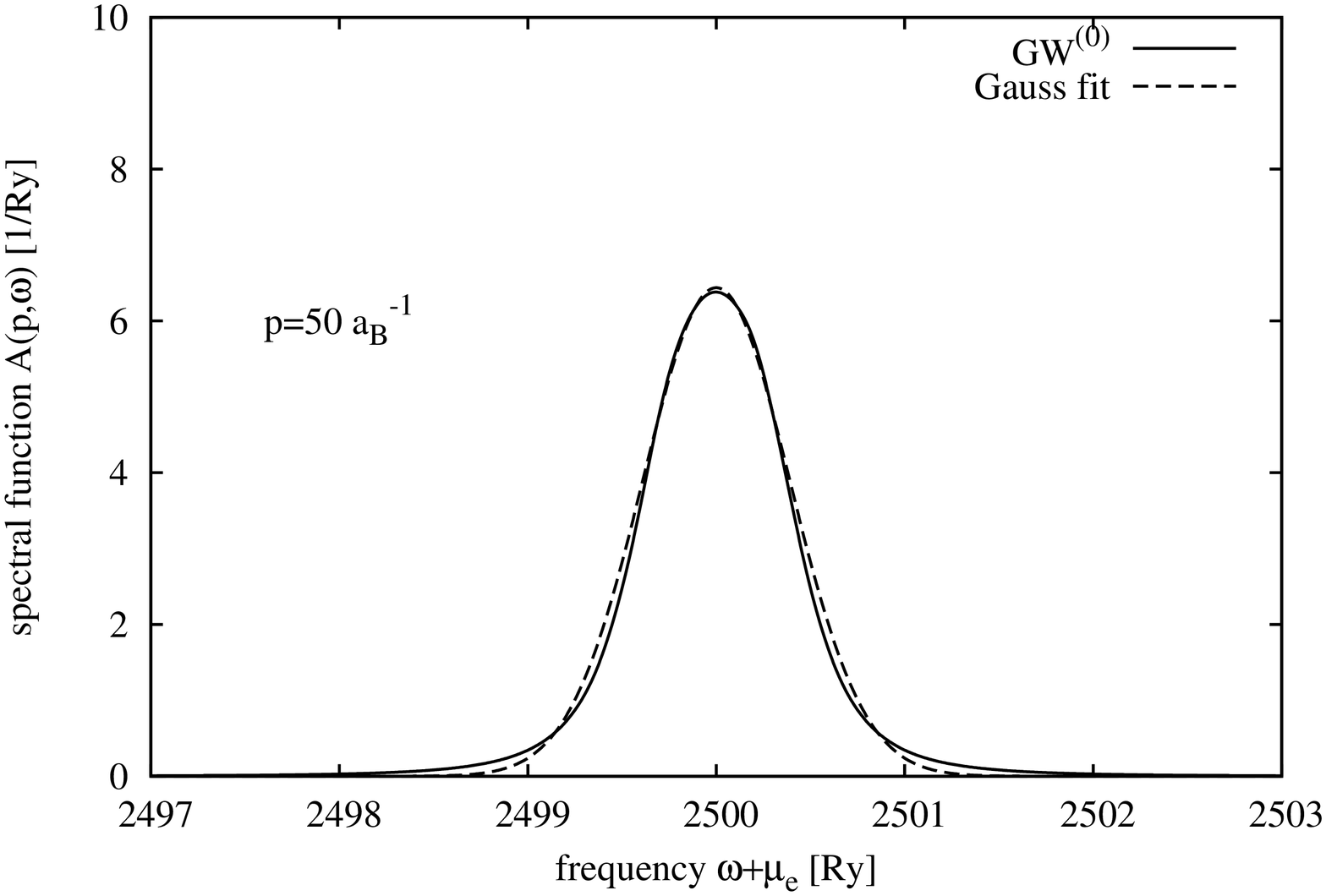}}%
    \subfigure[$p=100\,a_\mathrm{B}^{-1}$]{\includegraphics[width=.3\textwidth,angle=0,clip]{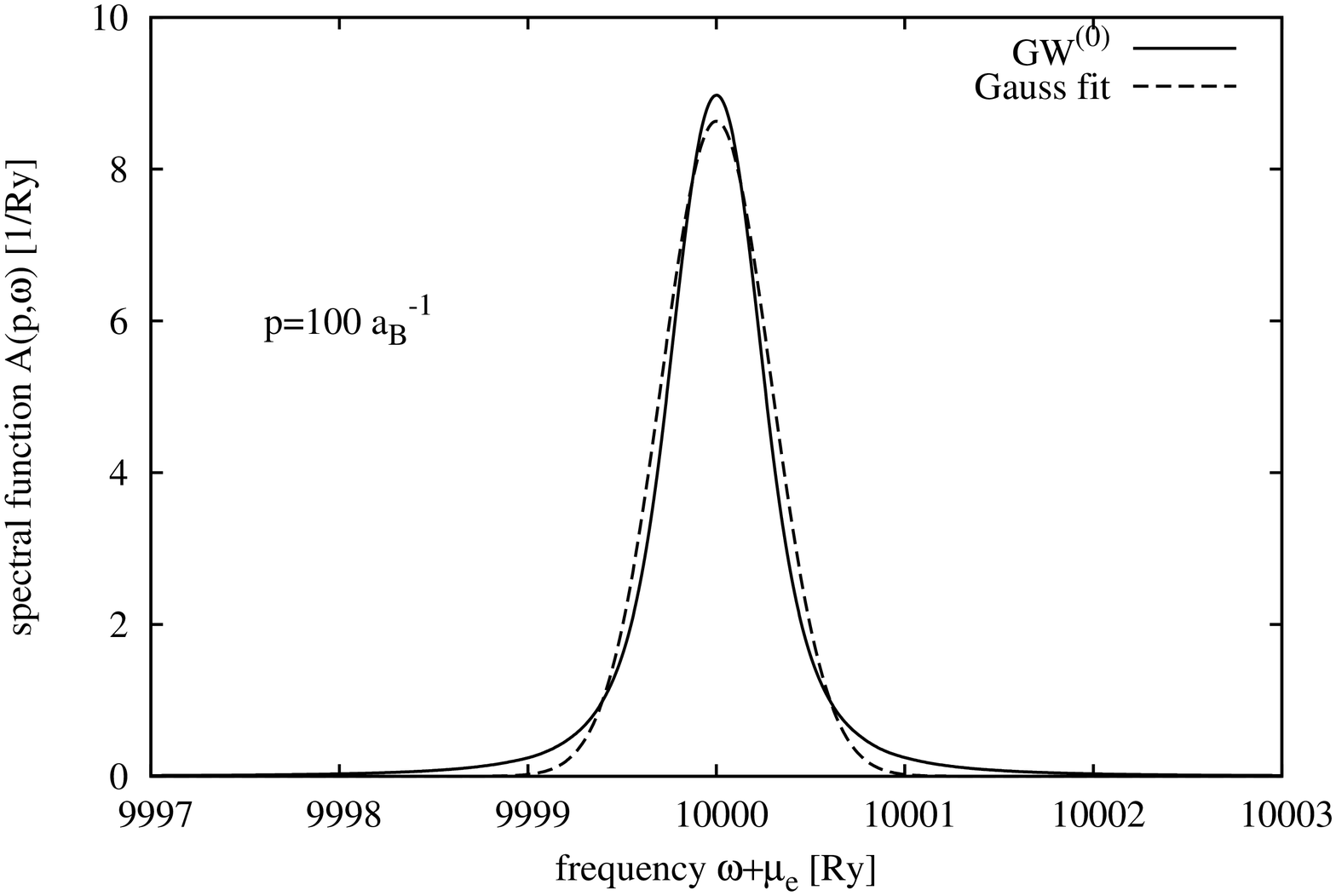}}%
  \end{center}
  \caption{Spectral function in $GW^{(0)}$-approximation (solid lines) and Gaussian ansatz  (dashed lines) with quasi-particle damping width $\sigma_p$ taken from
  equation (\ref{eqn:sigma_p_interpolation}) for three different momenta. Plasma
  parameters: $n=7\times 10^{20}\,\mathrm{cm^{-3}}$, $T=100\,\mathrm{eV}$. The plasma coupling parameter is
  $\Gamma=2.1\times 10^{-2}$, the degeneracy parameter is $\theta=3.5\times 10^{2}$, the Debye screening parameter is $\kappa=1.9\times
  10^{-2}\,a_\mathrm{B}^{-1}$.}
  \label{fig:sf_GW-Gauss_n7e20}
\end{figure}
\begin{figure}[ht]
  \begin{center}
    \subfigure[$p=0$]{\includegraphics[width=.3\textwidth,angle=0,clip]{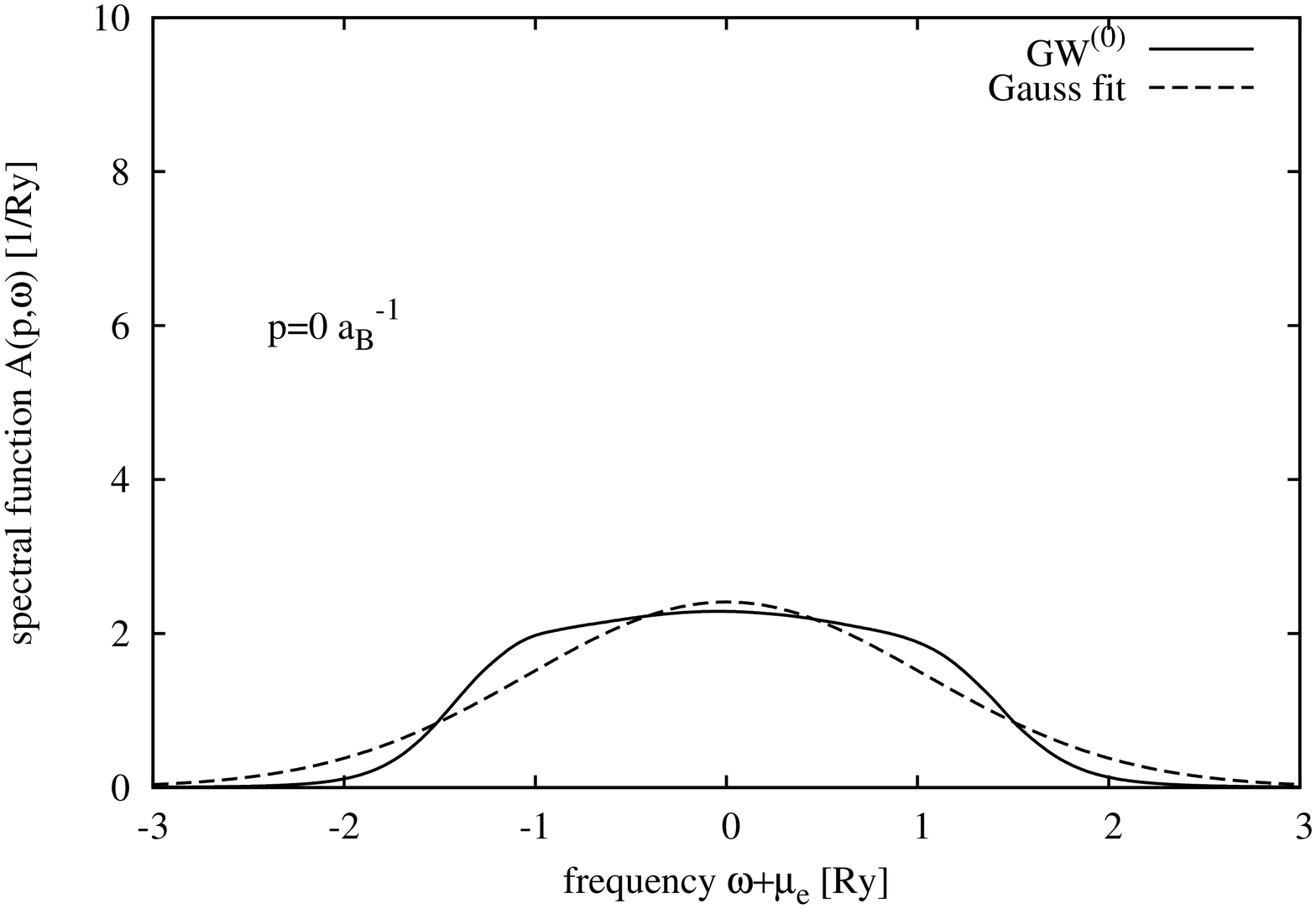}}%
    \subfigure[$p=50\,a_\mathrm{B}^{-1}$]{\includegraphics[width=.3\textwidth,angle=0,clip]{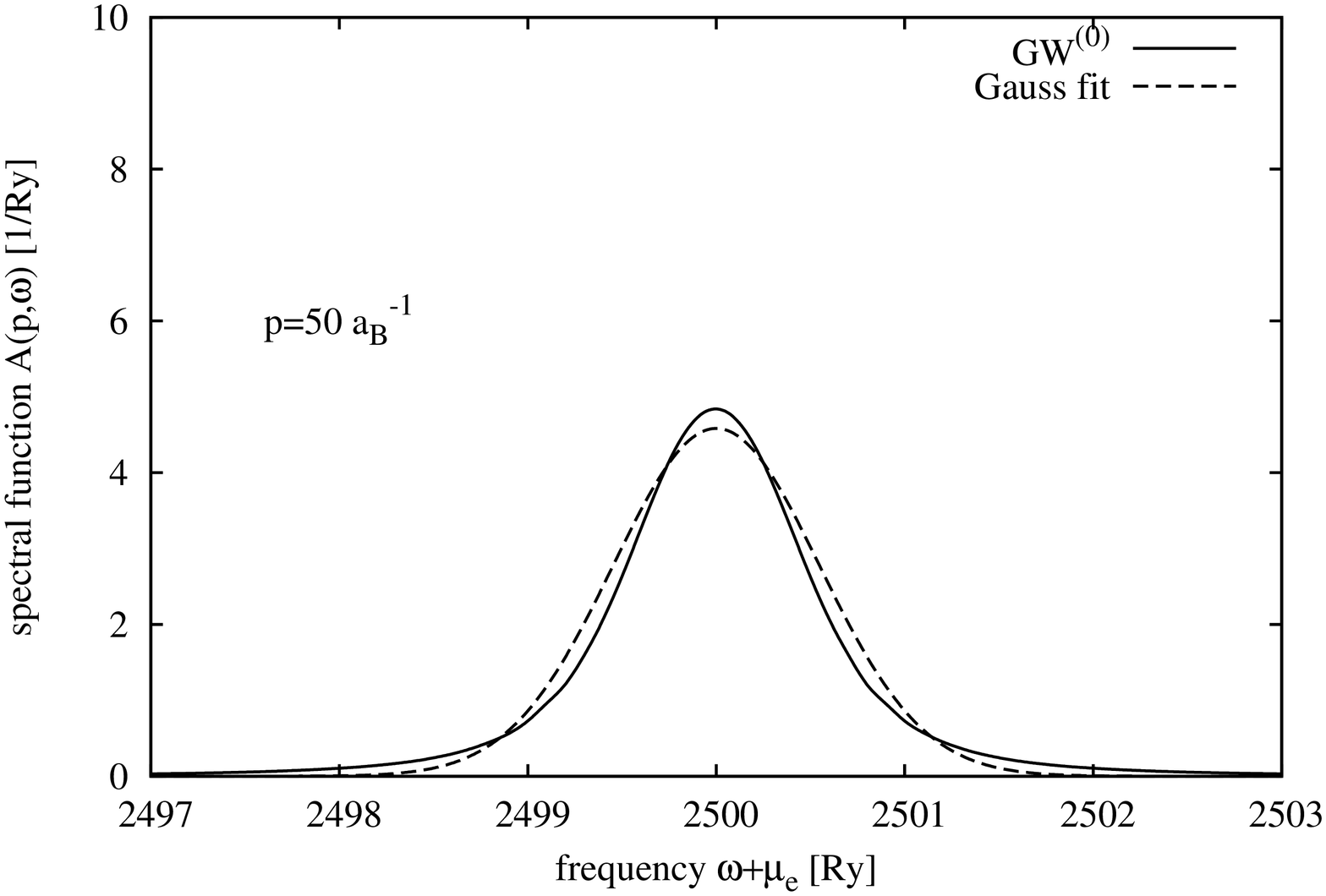}}%
    \subfigure[$p=100\,a_\mathrm{B}^{-1}$]{\includegraphics[width=.3\textwidth,angle=0,clip]{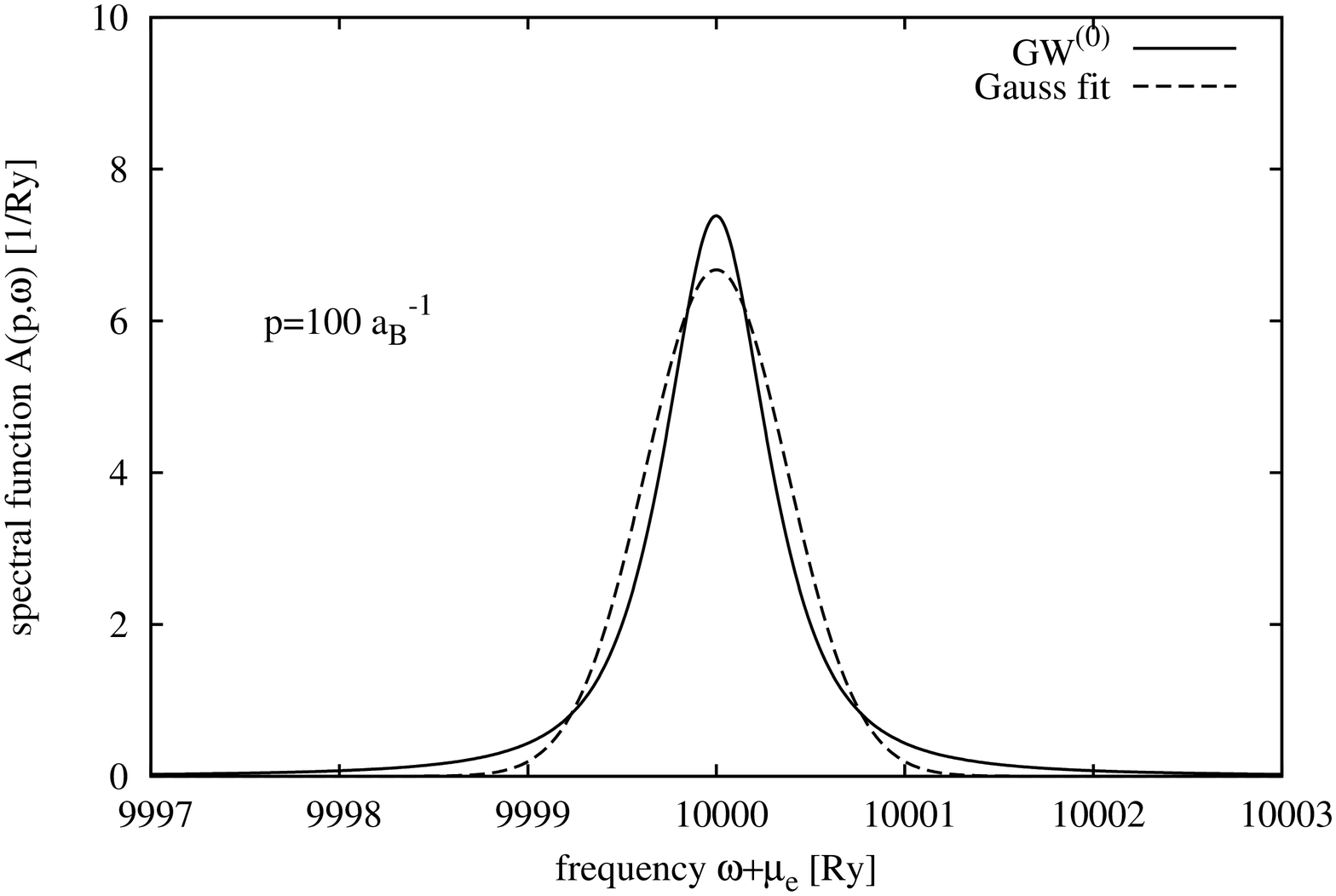}}%
  \end{center}
  \caption{Spectral function in $GW^{(0)}$-approximation (solid lines) and Gaussian ansatz  (dashed lines) with quasi-particle damping width $\sigma_p$ taken from
  equation (\ref{eqn:sigma_p_interpolation}) for three different momenta. Plasma
  parameters: $n=7\times 10^{21}\,\mathrm{cm^{-3}}$, $T=100\,\mathrm{eV}$. The plasma coupling parameter is
  $\Gamma=4.4\times 10^{-2}$, the degeneracy parameter is $\theta=7.5\times 10^{1}$, the Debye screening parameter is $\kappa=6.0\times
  10^{-2}\,a_\mathrm{B}^{-1}$.}
  \label{fig:sf_GW-Gauss_n7e21}
\end{figure}
\begin{figure}[ht]
  \begin{center}
    \subfigure[$p=0$]{\includegraphics[width=.3\textwidth,angle=0,clip]{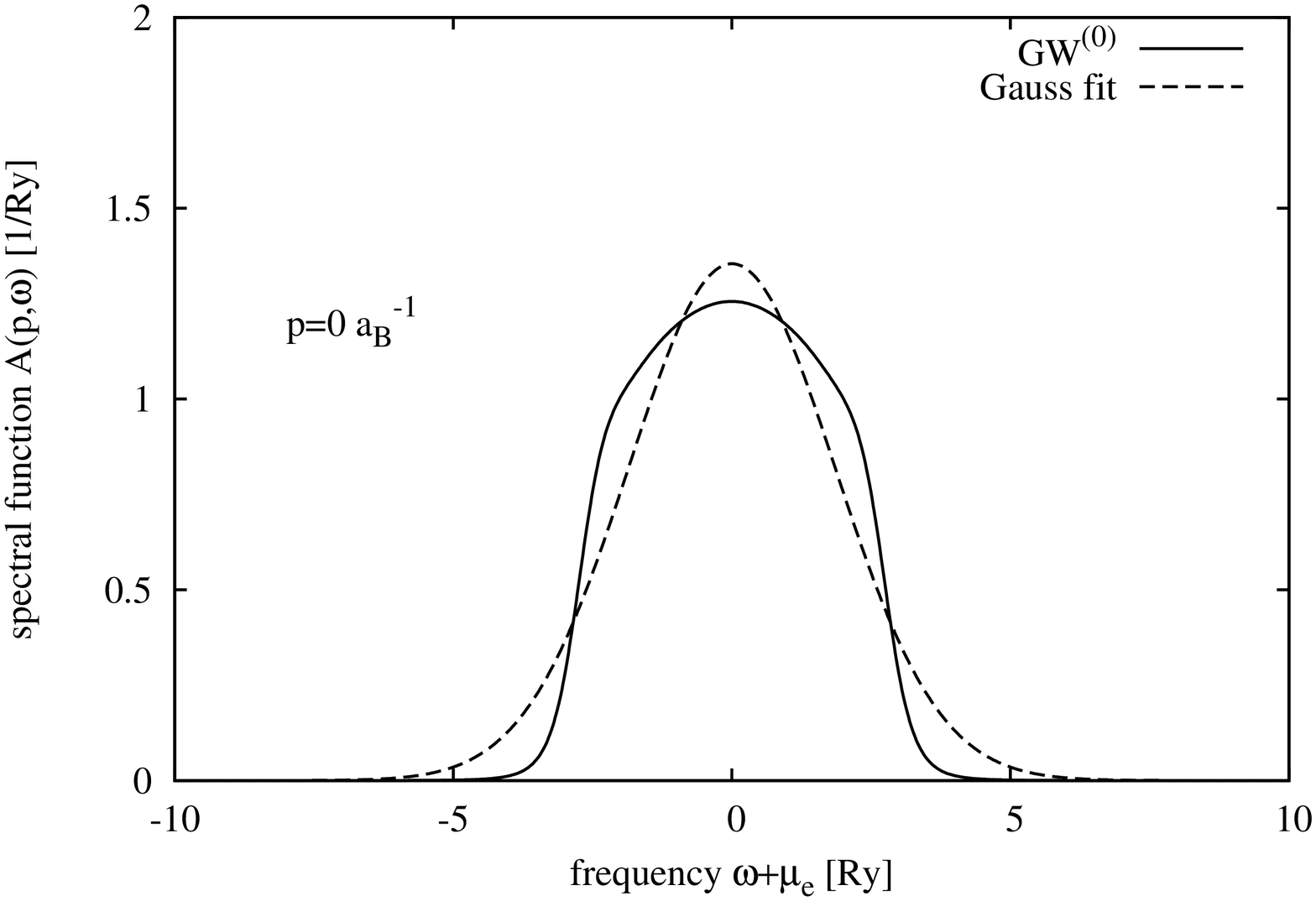}}%
    \subfigure[$p=50\,a_\mathrm{B}^{-1}$]{\includegraphics[width=.3\textwidth,angle=0,clip]{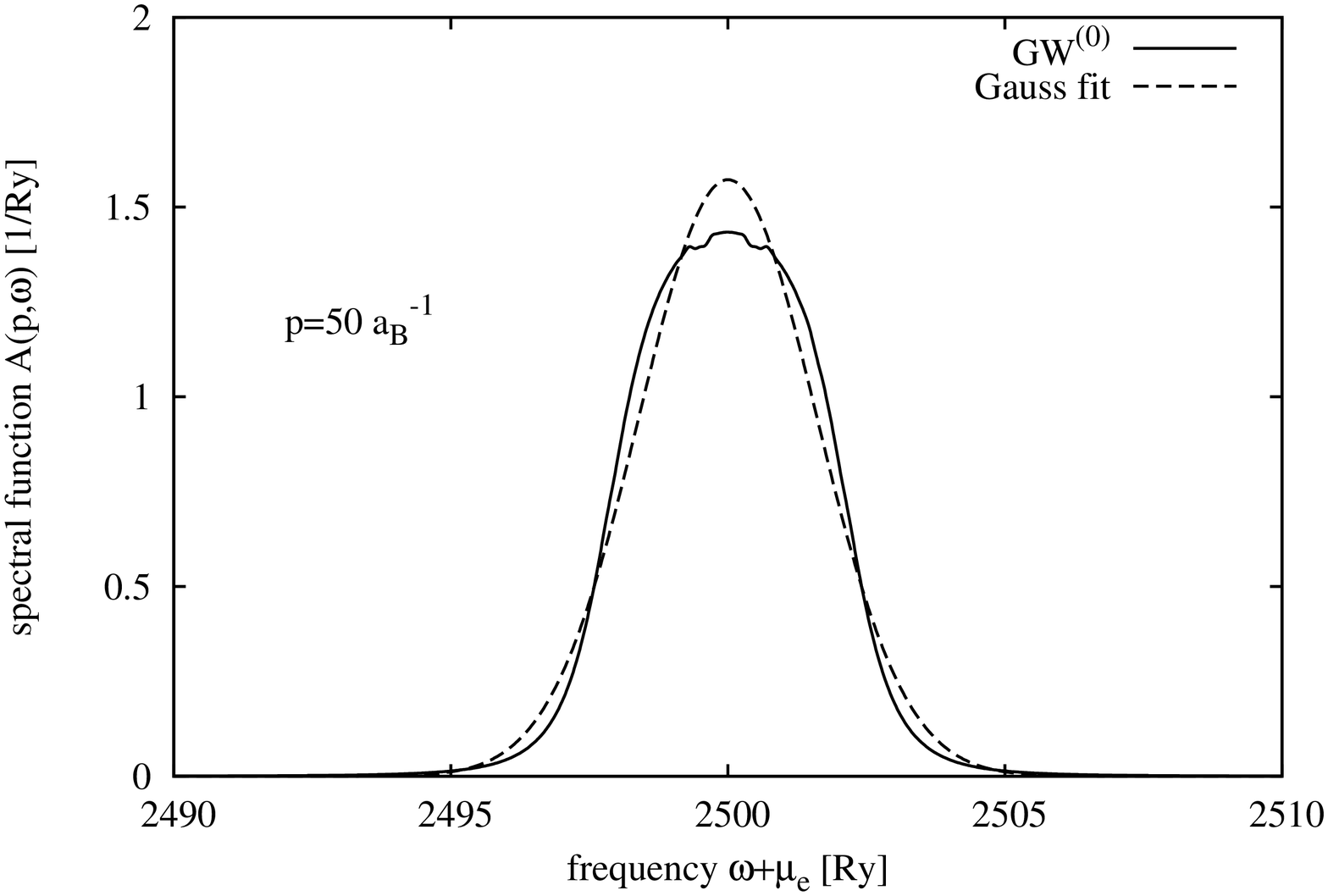}}%
    \subfigure[$p=100\,a_\mathrm{B}^{-1}$]{\includegraphics[width=.3\textwidth,angle=0,clip]{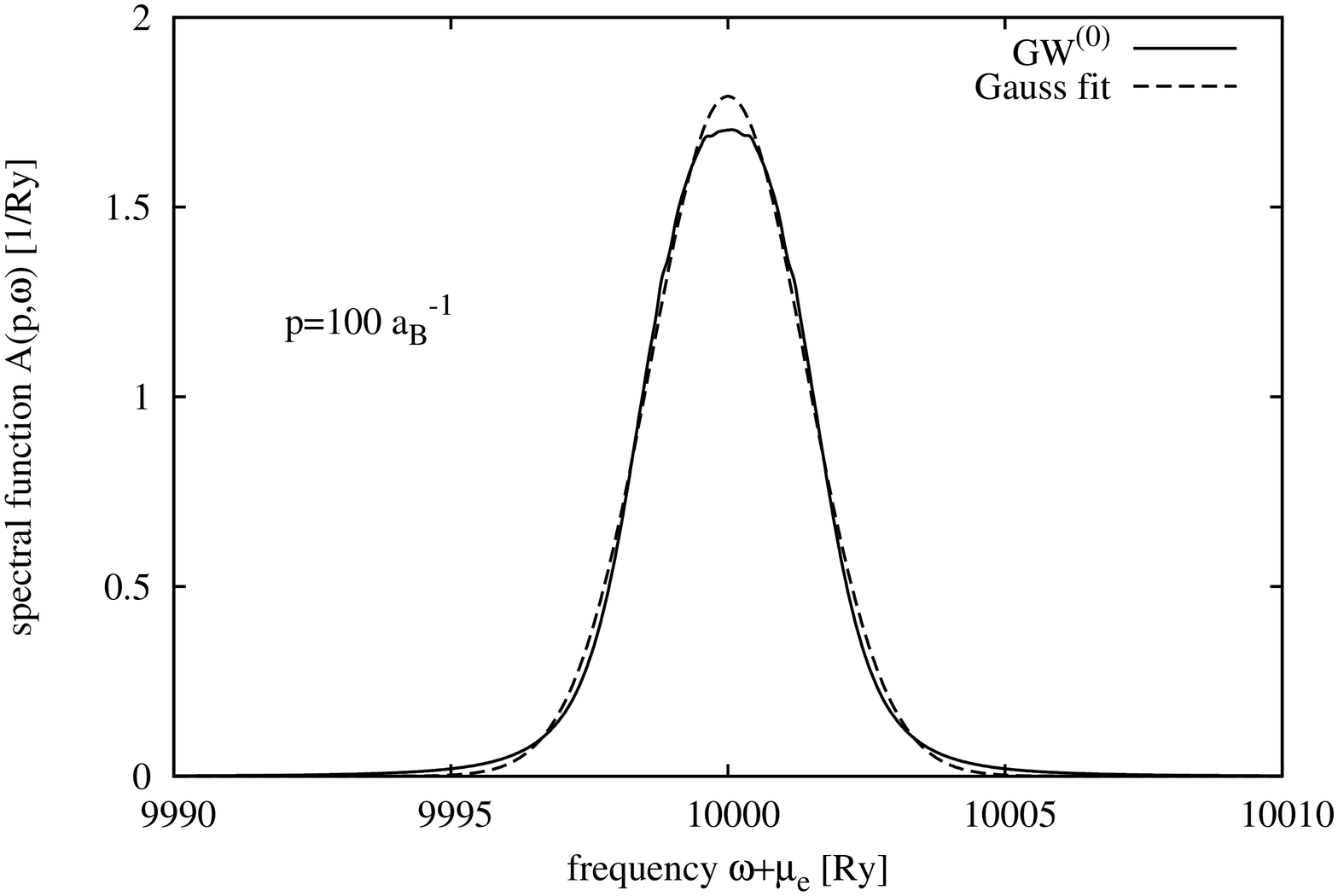}}%
  \end{center}
  \caption{Spectral function in $GW^{(0)}$-approximation (solid lines) and Gaussian ansatz  (dashed lines) with quasi-particle damping width $\sigma_p$ taken from
  equation (\ref{eqn:sigma_p_interpolation}) for three different momenta. Plasma
  parameters: $n=7\times 10^{21}\,\mathrm{cm^{-3}}$, $T=1000\,\mathrm{eV}$. The plasma coupling parameter is
  $\Gamma=4.4\times 10^{-3}$, the degeneracy parameter is $\theta=7.5\times 10^{2}$, the Debye screening parameter is $\kappa=1.9\times
  10^{-2}\,a_\mathrm{B}^{-1}$.}
  \label{fig:sf_GW-Gauss_n7e21_T1000}
\end{figure}
\begin{figure}[ht]
  \begin{center}
    \subfigure[$p=0$]{\includegraphics[width=.3\textwidth,angle=0,clip]{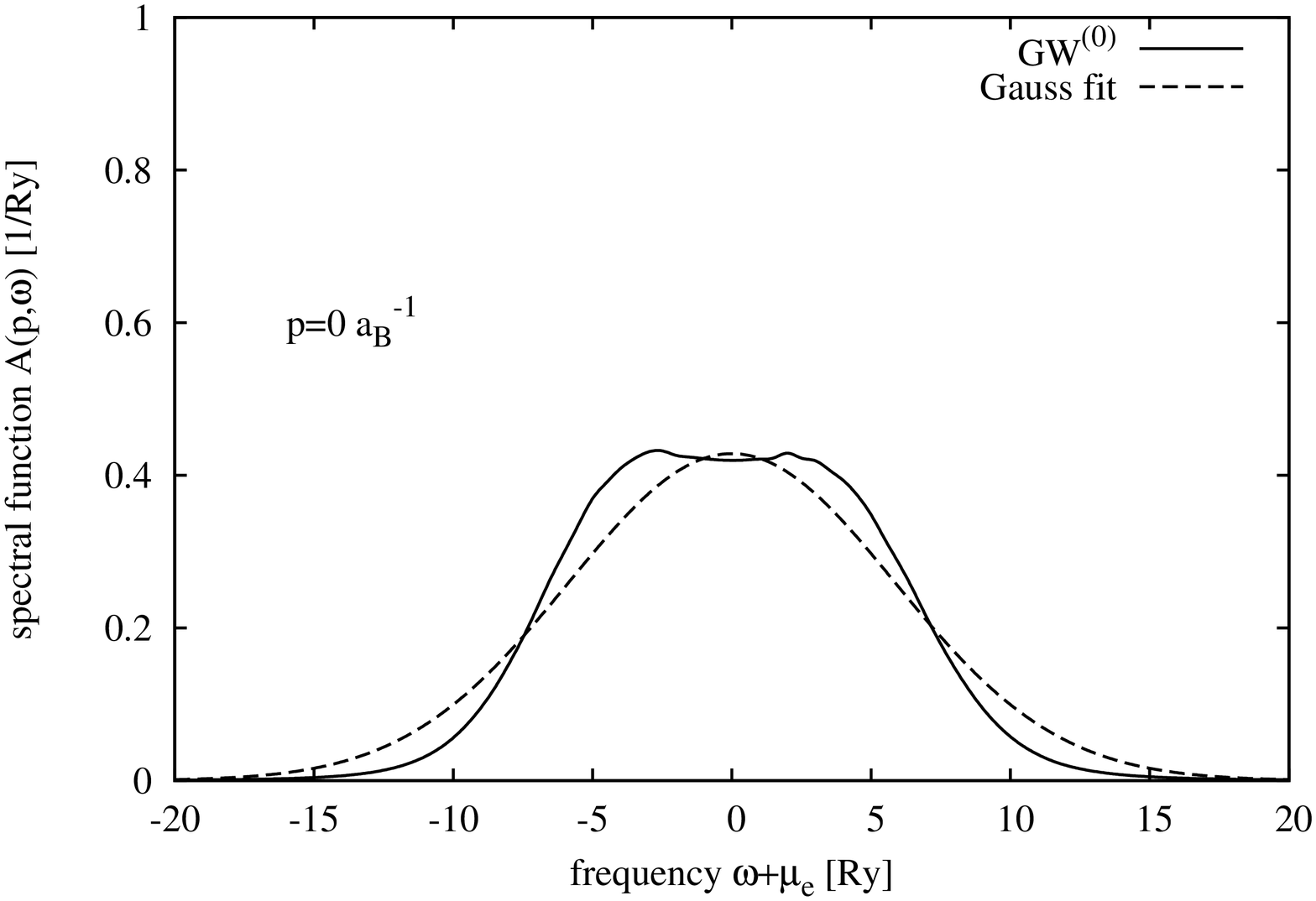}}%
    \subfigure[$p=50\,a_\mathrm{B}^{-1}$]{\includegraphics[width=.3\textwidth,angle=0,clip]{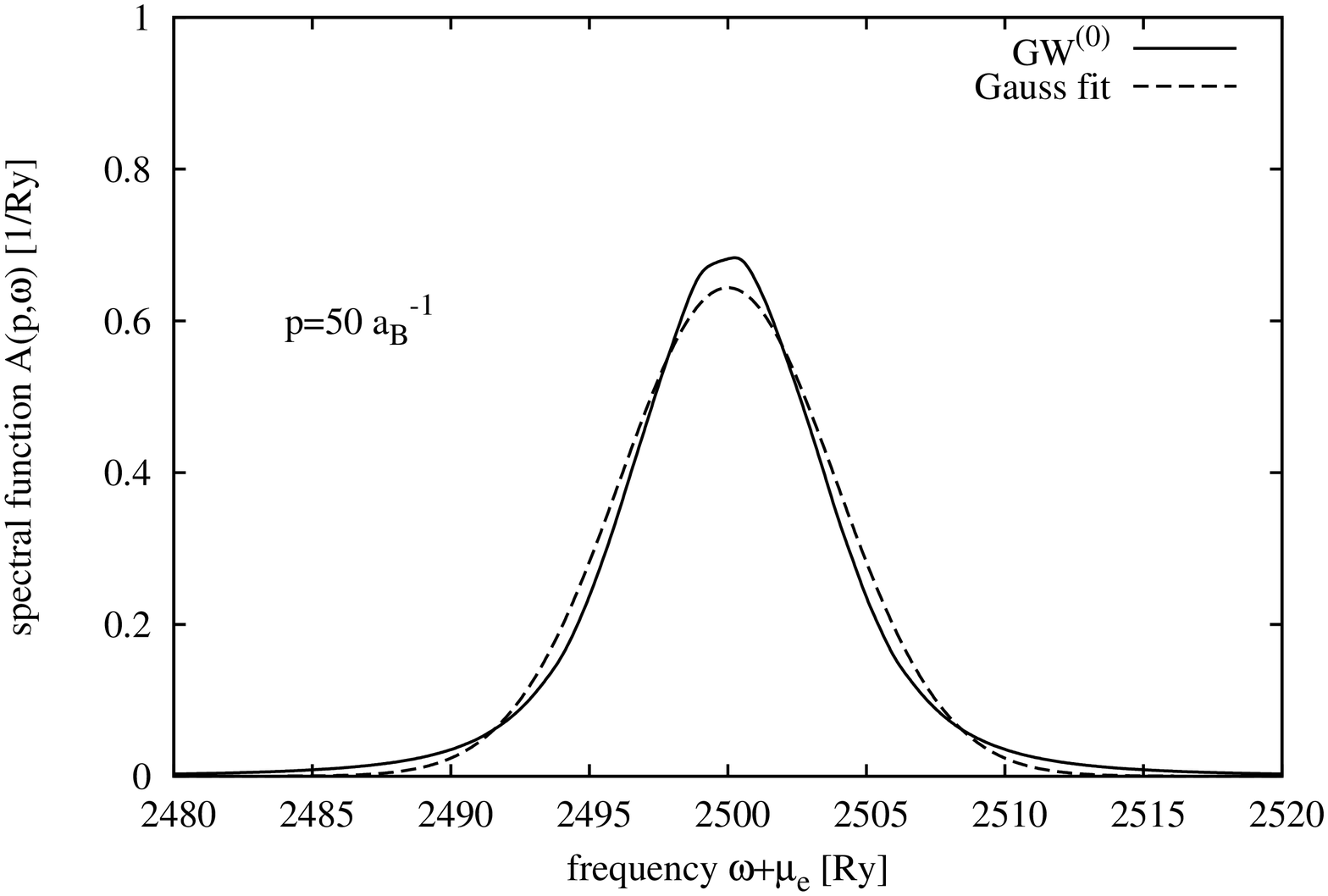}}%
    \subfigure[$p=100\,a_\mathrm{B}^{-1}$]{\includegraphics[width=.3\textwidth,angle=0,clip]{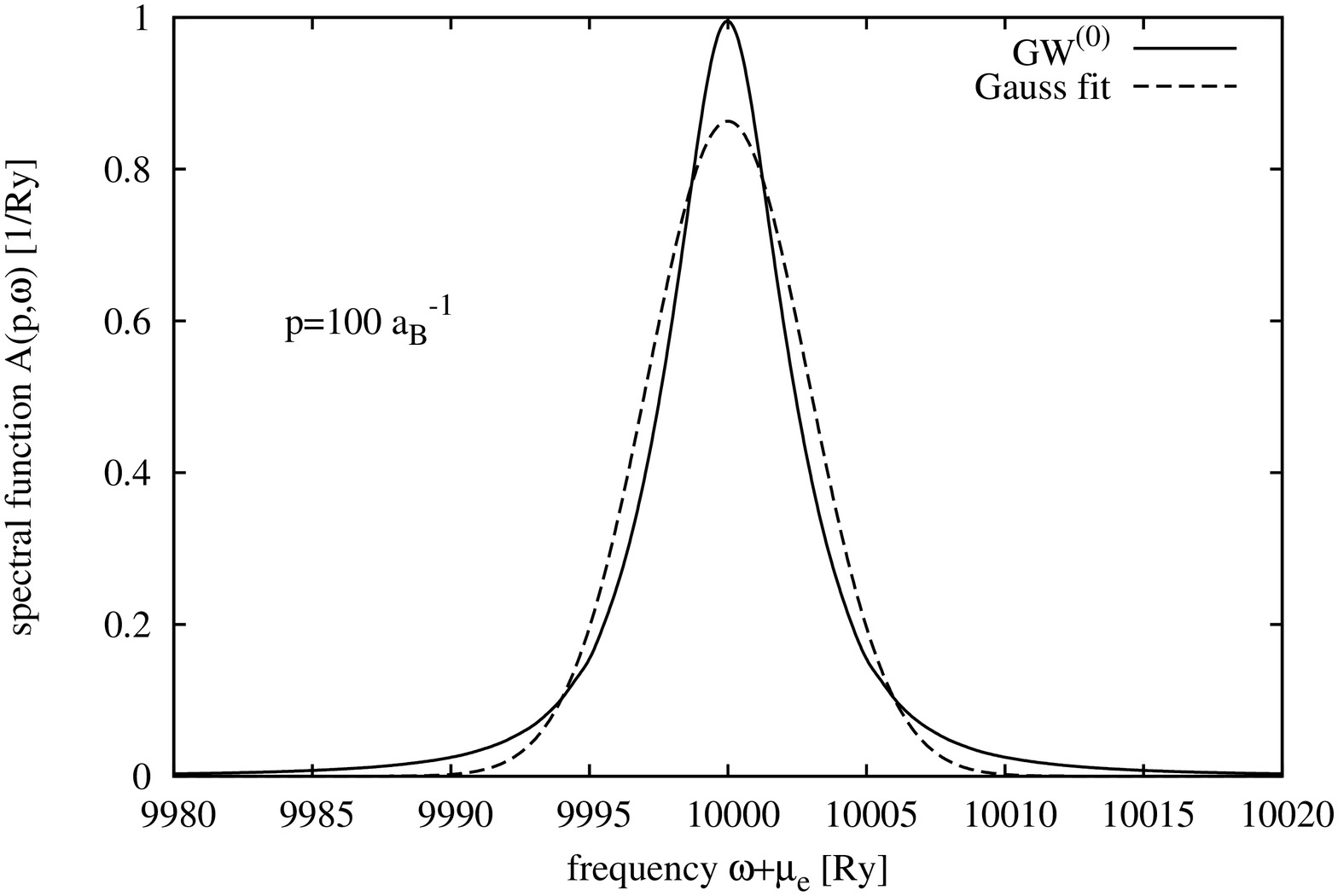}}%
  \end{center}
  \caption{Spectral function in $GW^{(0)}$-approximation (solid lines) and Gaussian ansatz  (dashed lines) with quasi-particle damping width $\sigma_p$ taken from
  equation (\ref{eqn:sigma_p_interpolation}) for three different momenta. Plasma
  parameters: $n=7\times 10^{23}\,\mathrm{cm^{-3}}$, $T=1000\,\mathrm{eV}$. The plasma coupling parameter is
  $\Gamma=2.1\times 10^{-2}$, the degeneracy parameter is $\theta=3.5\times 10^{1}$, the Debye screening parameter is $\kappa=1.9\times
  10^{-1}\,a_\mathrm{B}^{-1}$.}
  \label{fig:sf_GW-Gauss_n7e23_T1000}
\end{figure}
\begin{figure}[ht]
  \begin{center}
    \subfigure[$p=0$]{\includegraphics[width=.3\textwidth,angle=0,clip]{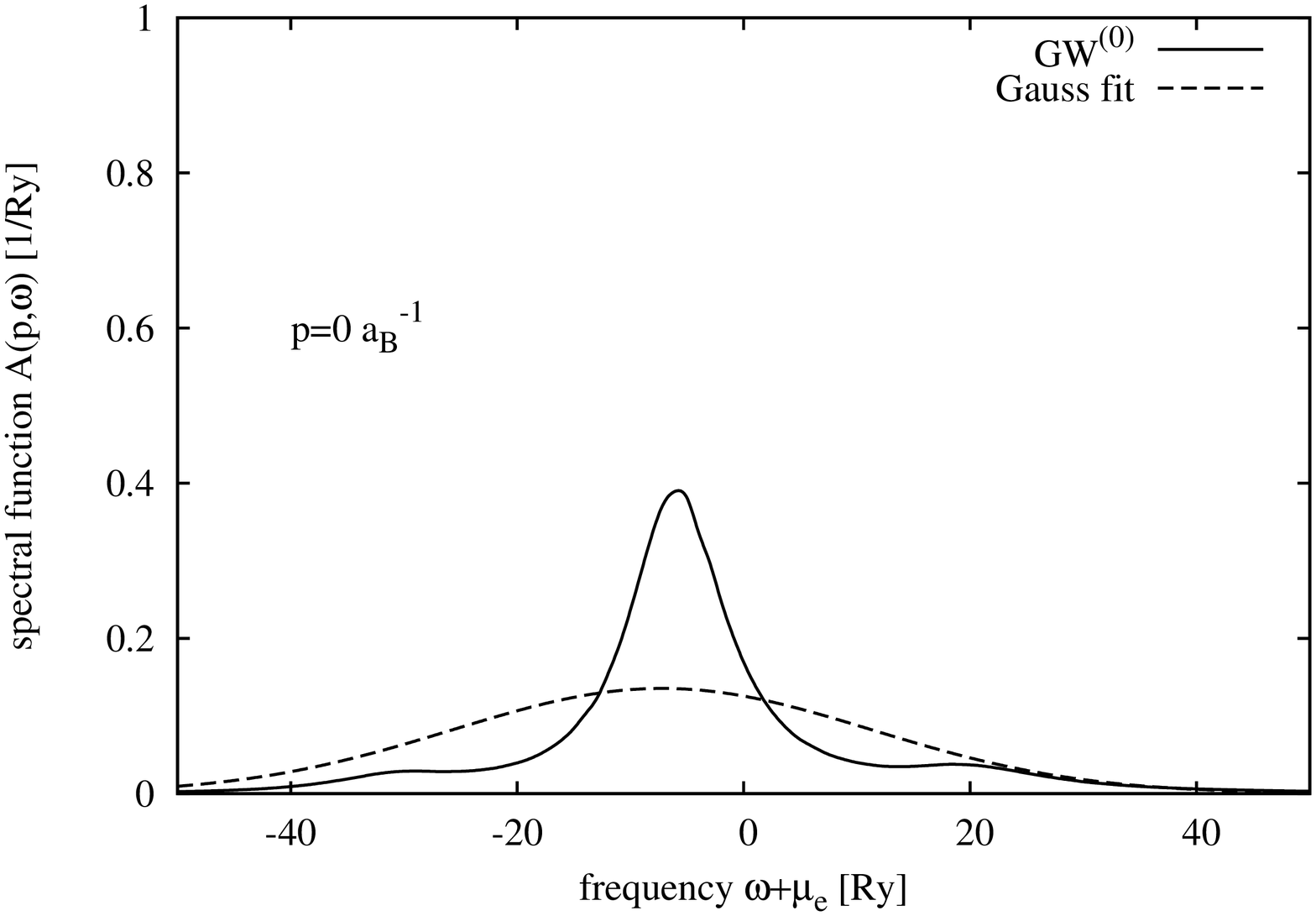}}%
    \subfigure[$p=50\,a_\mathrm{B}^{-1}$]{\includegraphics[width=.3\textwidth,angle=0,clip]{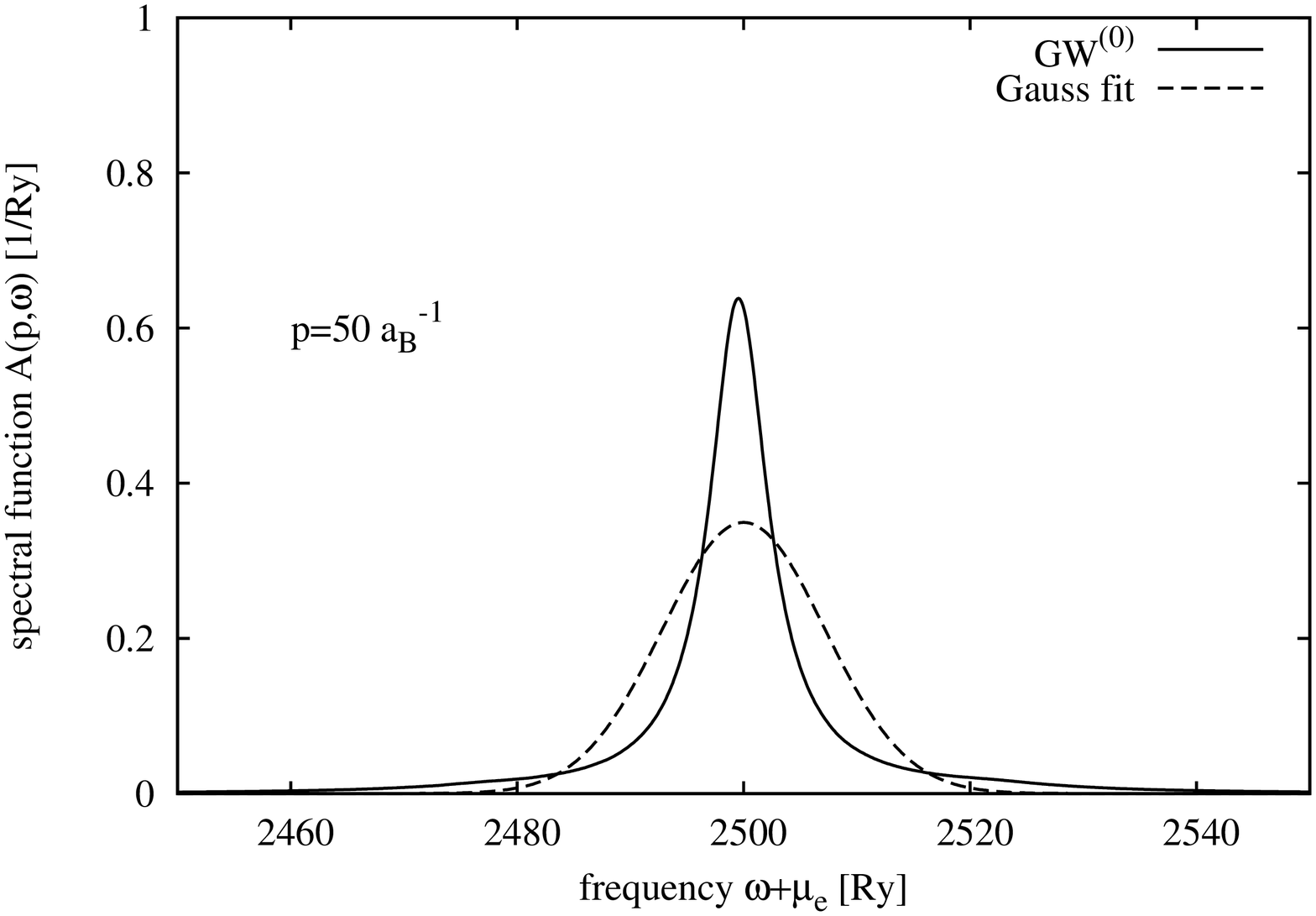}}%
    \subfigure[$p=100\,a_\mathrm{B}^{-1}$]{\includegraphics[width=.3\textwidth,angle=0,clip]{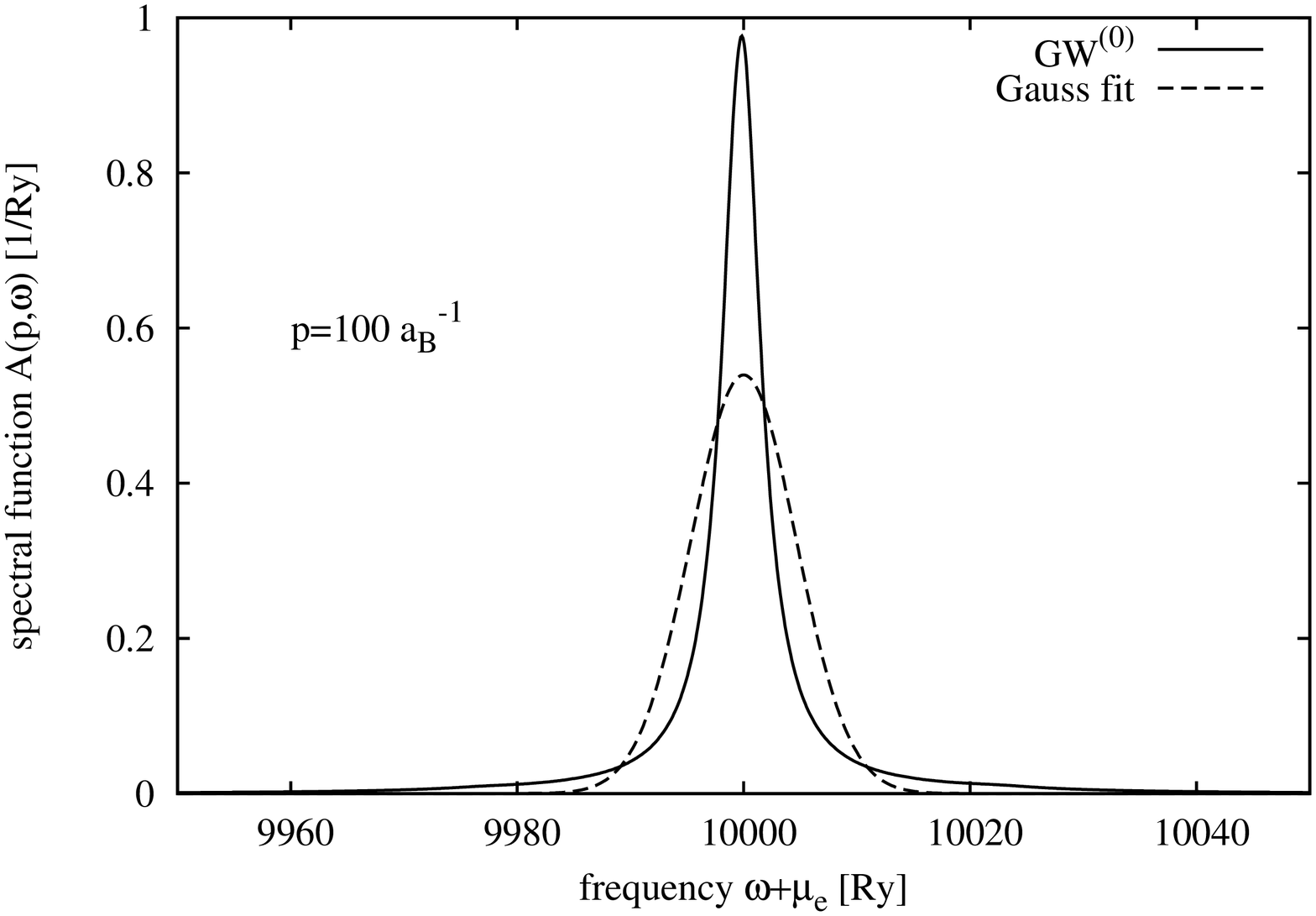}}%
  \end{center}
  \caption{Spectral function in $GW^{(0)}$-approximation (solid lines) and Gaussian ansatz  (dashed lines) with quasi-particle damping width $\sigma_p$ taken from
  equation (\ref{eqn:sigma_p_interpolation}) for three different momenta. Plasma
  parameters: $n=7\times 10^{25}\,\mathrm{cm^{-3}}$, $T=1000\,\mathrm{eV}$. The plasma coupling parameter is
  $\Gamma=9.6\times 10^{-2}$, the degeneracy parameter is $\theta=1.6$, the Debye screening parameter is $\kappa=1.9\,a_\mathrm{B}^{-1}$. Here,
  the Gaussian fit is no longer sufficient due to the appearance of plasmaron resonances in the spectral function (shoulders at $\omega\simeq
  -30\,\mathrm{Ry}$ and $\omega\simeq 20\,\mathrm{Ry}$).}
  \label{fig:sf_GW-Gauss_n7e25_T1000}
\end{figure}

At smaller densities, the correspondence is even
better as can be seen by comparing the spectral functions shown in figures \ref{fig:sf_GW-Gauss_n7e19} - \ref{fig:sf_GW-Gauss_n7e25_T1000}.
The dashed curves give the Gaussian ansatz for the spectral function 
with the quasi-particle width taken from the interpolation formula~(\ref{eqn:sigma_p_interpolation}).
As a general result, the analytic expression for the quasi-particle damping $\sigma_p$ leads to a spectral function that nicely fits the numerical
solution for the spectral function at least at finite $p$. At very small values of $p$, the overall correspondence is still fair, i.e. the
position of the maximum and the overall width match, but the detailed behaviour does not coincide. In particular, the steep wings and the
central plateau, that forms in the
$GW^{(0)}$-calculation, is not reproduced by the one-parameter Gaussian. 
For this situation, the analytic formula for self-energy  given in \cite{Fortmann_JPhysA41_445501_2008} should be used instead.

By comparing the numerical data for the spectral function to the Gaussian ansatz at different densities, it is found, that the
Gaussian spectral function is a good approximation as long as the Debye screening parameter $\kappa$ is smaller than the inverse Bohr radius, 
$\kappa<1\,a_\mathrm{B}^{-1}$.
This becomes immanent by comparing figures \ref{fig:sf_GW-Gauss_n7e23_T1000} and \ref{fig:sf_GW-Gauss_n7e25_T1000}. In the first case
($n_\mathrm{e}=7\times 10^{23}\,\mathrm{cm}^{-3}, T=1000\,\mathrm{eV}$), we have $\kappa=0.19$, while in the second case
($n_\mathrm{e}=7\times 10^{25}\,\mathrm{cm}^{-3}, T=1000\,\mathrm{eV}$), $\kappa=1.9$ is found. As already noted in the discussion of the
numerical results in section \ref{sec:numresults}, in the case of increased density, the plasmaron satellites appear as separate structures in the
wings of the central quasi-particle peak, whereas they are hidden in the central peak at smaller densities. Therefore, a single Gaussian
is not sufficient to fit the spectral function at increased densities.
Since the position of the plasmaron peak is given approximatively by the plasma frequency $\omega_\mathrm{pl}$, whereas the width of the central
peak at small $p$ is 
just the quasi-particle width $\sigma_0$, we can identify the ratio of these two quantities, $-\omega_\mathrm{pl}/\sigma_0\propto
\sqrt{\kappa}$ as
the parameter which tells us if plasmaron peaks appear separately ($\omega_\mathrm{pl}>-\sigma_0$) or not ($\omega_\mathrm{pl}<-\sigma_0$).
Since the plasma frequency increases as a function of $n_\mathrm{e}^{1/2}$, whereas the quasi-particle width scales as 
$n^{1/4}$ (c.f. equation (\ref{eqn:sigma_p_interpolation})), the transition from the single peak behaviour to the more complex behaviour including plasmaron resonances,
appears at increased density. Neglecting numerical constants of order 1 in the ratio of plasma frequency to damping width, 
we see that $-\omega_\mathrm{pl}/\sigma_0<1$ is
equivalent to $\kappa<1$, which was our observation from the numerical results. Therefore, we can identify the range of validity of the presented 
expressions for the spectral function and the  quasi-particle damping. It is valid for those plasmas, where we have densities and temperatures,
such that $\kappa<1$.
 
The physical origin of the requirement $\kappa<1$ can be understood in the following way \cite{Fortmann_JPhysA41_445501_2008}.
At length scales smaller than the Bohr radius, one typically expects quantum effects, e.g. Pauli blocking. These effects are
not accounted for in the derivation of the quasi-particle damping. 
Therefore, it appears to be a logical consequence that the validity of the
results is limited by the length scale at which typical quantum phenomena become important.

The regime of validity of the analytic formula can also be expressed via the plasma coupling parameter and the temperature
as $\Gamma <T^{-2/3}.$ Since we restrict
ourselves to plasma temperatures, where bound states can be excluded, i.e. $T>1\,\mathrm{Ry}$, this is equivalent to saying $\Gamma<1$.

Although the correspondence between the accurate $GW^{(0)}$ calculations and the parametrized spectral function at small momenta 
is not as good as in the case of large momenta,  the parametrized  spectral function can be applied in the regime of validity to the calculation of plasma observables without introducing too large errors.
As an example, this will be shown for the case of the chemical potential $\mu$ in the next section.

\section{Application: Shift of the chemical potential\label{sec:chempot}}
To demonstrate the applicability of the presented formulae for quick and reliable calculations of plasma properties, we calculate the shift of the 
electron's chemical potential $\Delta\mu=\mu-\mu_\mathrm{free}$, i.e. the deviation of the chemical potential of the interacting
plasma $\mu$ from the value of the non-interacting system $\mu_\mathrm{free}$. The chemical potential of the interacting system
$\mu$ is obtained by inversion of the density as a function of $T$ and $\mu$, equation
(\ref{eqn:densityrelation}). The free chemical potential $\mu_\mathrm{free}$ is obtained in a similar way by inversion of the free density
\begin{equation}
  n_\mathrm{free}(T,\mu_\mathrm{free})=2\sum_{\mathbf{p}}n_\mathrm{F}(\varepsilon_\mathbf{p}-\mu_\mathrm{free})~.
  \label{eqn:densityrelationfree}
\end{equation}

\begin{figure}[ht]
  \begin{center}
    \includegraphics[width=.5\textwidth,angle=0,clip]{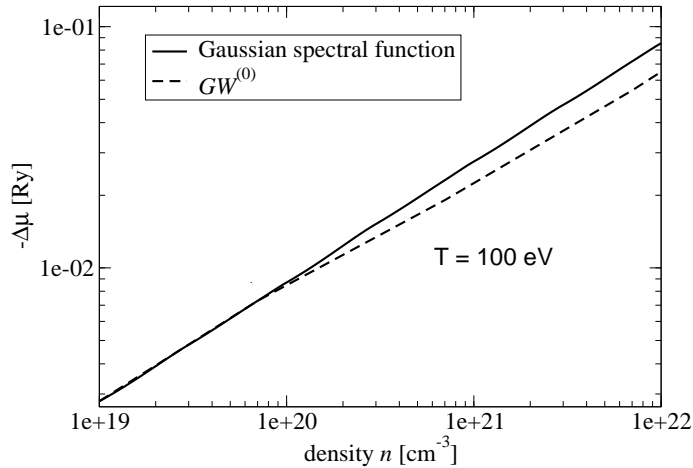}
  \end{center}
  \caption{Shift of the chemical potential as a function of the plasma density for a plasma temperature $T=100\,\mathrm{eV}$. Results for
  $\Delta\mu$ using the parametrized spectral function (solid line) are compared to full numerical calculations, using the
  $GW^{(0)}$-approximation. }
  \label{fig:chempotshift}
\end{figure}

Figure \ref{fig:chempotshift} shows the shift of the chemical potential as a function of the plasma 
density $n$ for a fixed plasma temperature $T=100\,\mathrm{eV}$. Results
obtained by inversion of equation~(\ref{eqn:densityrelation}) using the parametrized spectral function (\ref{eqn:GaussDef}) with the
quasi-particle damping width taken from equation~(\ref{eqn:sigma_p_interpolation}) (solid curve) are compared to those results taking the numerical
$GW^{(0)}$ spectral function (dashed curve).

The $GW^{(0)}$ result gives slightly smaller shifts than the parametrized spectral function, i.e. the usage of the analytical damping width leads
to an overestimation of the the shift of the chemical potential. However, the deviation 
remains smaller than 20\% over the range of densities considered here, i.e. for $\kappa<1$. At small densities, i.e. for $n\leq
10^{20}\,\mathrm{cm}^{-3}$, the parametrized spectral function yields the same result as the full $GW^{(0)}$ calculation.

The deviation at increased density can be reduced by improving the parametrization of the spectral function at small momenta. To this end, the behaviour of the
quasi-particle damping width at high momenta, equation (\ref{eqn:finalsolution}) should be combined with the frequency dependent solution for
$\sigma_p$ at vanishing momentum, as presented in Ref.~\cite{Fortmann_JPhysA41_445501_2008}. However, this task goes beyond the scope of this paper,
where we wish to present comparatively simple analytic expressions for the damping width that yield the correct low density behaviour of plasma
properties.
 
\section{Conclusions and Outlook}
In this paper, the $GW^{(0)}$-approximation for the single-particle self-energy was evaluated for the case of a classical one-component electron
plasma, with ions treated as a homogeneous charge background. A systematic behaviour of the spectral function was found, i.e. a symmetrically
broadened structure at low momenta and convergence to a sharp quasi-particle resonance at high $p$. At increased densities, plasmaron satellites
show up in the spectral function as satellites besides the main peak. 

In the second part, an analytic formula for the imaginary part of the self-energy at the quasi-particle dispersion
$\omega^\mathrm{QP}(\mathbf{p})=\varepsilon_\mathbf{p}+\Sigma^\mathrm{HF}(\mathbf{p})$
was derived as a two-point Pad\'e formula that interpolates between the exactly known behaviour
at $p=0$ and $p\to \infty$. The former case was studied in \cite{Fortmann_JPhysA41_445501_2008}, while an expression for the asymptotic case
$p\to \infty$ was derived here. The result is summarized in equation (\ref{eqn:sigma_p_interpolation}). In contrast to previously known expressions
for the quasi-particle damping, based on a perturbative approach to the self-energy \cite{FennelWilfer_AnnPhysL32_265_1974}, the result presented
here shows a physically intuitive behaviour in the limit of low densities, i.e. it vanishes when the system becomes dilute.
Using the Gaussian ansatz (\ref{eqn:GaussDef}) for the spectral function in combination with the quasi-particle width leads to a very good
agreement with the numerical data for the spectral function in the range of plasma parameters, where $\kappa<1\,a_\mathrm{B}^{-1}$; the relative 
deviation is smaller than 10\% under this constraint.

Thus, a simple expression for the damping width of electrons in a classical plasma has been found, that can be used to approximate the full 
spectral function to high accuracy. This achievement greatly facilitates the calculation of observables that take the spectral function or the self-energy 
as an input, such as optical properties (inverse bremsstrahlung absorption), conductivity, or the stopping power. 

Furthermore, it was demonstrated that the derived expressions allow for quick and reliable calculations of plasma properties without having to
resort to the full self-consistent solution of the $GW^{(0)}$-approximation. As an example, the shift of the chemical potential was calculated
using the parametrized spectral function and compared to $GW^{(0)}$ results.
For densities of $n<10^{21}\,\mathrm{cm}^{-3}$, 
both approaches coincide with a relative deviation of less than 10\%, 
going eventually up to 20\% as the density approaches
$10^{22}\,\mathrm{cm^{-3}}$.
At low densities both approaches give identical results.
This shows the extreme usefulness of the presented approach for the calculation of observables via the parametrized spectral function.

As a furhter important application of the results presented in this paper, we would like to mention the calculation of radiative energy loss of particles
traversing a dense medium, i.e. bremsstrahlung. A many-body theoretical approach to this scenario is 
given by Knoll and Voskresensky \cite{knol:annals96}, using non-equilibrium Green functions. 
They showed, that a finite spectral width of
the emitting particles leads to a decrease in the bremsstrahlung emission. This effect is known as the Landau-Pomeranchuk-Migdal effect
\cite{migd:physrev56,klei:revmodphys99}. It has been experimentally confirmed in relativistic electron scattering experiments using dense targets,
e.g. lead \cite{hans:prl03,hans:prd04}. 
In \cite{Fortmann:CMT20_2007}, it is shown that also thermal bremsstrahlung from a plasma 
is reduced due to the finite spectral width of the electrons in the plasma. In the cited papers, the quasi-particle damping width was either set as a momentum- and energy
independant parameter (in \cite{knol:annals96}), or calculated self-consistently using simplified approximations of the $GW^{(0)}$ theory 
(in \cite{Fortmann:CMT20_2007}), which itself is a very time-consuming task and prohibited investigations over a broad range of plasmas
parameters. Now, based on this work's results, calculations on the level of full $GW^{(0)}$ become feasible, since analytic formulae have been found that reproduce the
$GW^{(0)}$ self-energy. Effects of dynamical correlations on the bremsstrahlung spectrum can be studied starting from a consistent single-particle
description via the $GW^{(0)}$ self-energy.

\begin{acknowledgments}
  The author acknowledges many helpful advice from Gerd R\"opke and fruitful discussion with W.-D. Kraeft as well as with C. D. Roberts.
  Financial support was obtained from the German Research Society (DFG) via the Collaborative Research Center ``Strong Correlations and Collective
  Effects in Radiation Fields: Coulomb Systems, Clusters, and Particles'' (SFB 652).
\end{acknowledgments}
%%%%%%%%%%%%%%%%%%%%%%%%%%%%%%%%%%%%%%%%%%%%%%%%%%%%%%%%%%%%%%%%%%%%%%%%%%%%%%%5
\begin{appendix}

\section{Analytic solution for the $GW^{(0)}$ self-energy using the plasmon pole approximation\label{app:GaussPPA}}

  After the angular integration which was performed in equation (\ref{eqn:thetaIntegration}), the imaginary part of the self-energy at the
  quasi-particle dispersion $\omega=\varepsilon_\mathbf{p}$ reads
\begin{multline}
  \mathrm{Im}\,\Sigma(\mathbf{p},p^2)=\frac{\omega_\mathrm{pl}^2}{4p}\int_{0}^{\infty}\frac{dq}{q\,\omega_\mathbf{q}}\Bigg\{ 
  \left[
  \text{Erf}\,\left( \frac{q^2+2pq+\omega_\mathbf{q}}{\sqrt{2}\sigma_p} \right)-
  \text{Erf}\,\left( \frac{q^2-2pq+\omega_\mathbf{q}}{\sqrt{2}\sigma_p} \right)
  \right]\,n_\mathrm{B}(\omega_\mathbf{q})\,\exp(\omega_\mathbf{q}/T)\\
  -\left[
  \text{Erf}\,\left( \frac{q^2+2pq-\omega_\mathbf{q}}{\sqrt{2}\sigma_p} \right)-
  \text{Erf}\,\left( \frac{q^2-2pq-\omega_\mathbf{q}}{\sqrt{2}\sigma_p} \right)
  \right]\,n_\mathrm{B}(-\omega_\mathbf{q})\,\exp(-\omega_\mathbf{q}/T)
  \Bigg\}~.
  \label{eqn:deriv0040}
\end{multline}

This equation represents a self-consistent equation for 
$\mathrm{Im}\,\Sigma(\mathbf{p},\omega=p^2)=\sqrt{2/\pi}\sigma_p$. 

Our aim is to derive an analytic expression, that approximates the numerical solution of equation~(\ref{eqn:deriv0040}) for arbitrary $p$.
To this end, we first look at the case of large momenta, $p\gg\kappa$, and later combine that result with known expressions for the limit of
vanishing momentum $p\to 0$, to produce an interpolation (``Pad\'e'') formula that covers the complete $p$-range. 

We perform a sequence of approximations to the integral in (\ref{eqn:deriv0040}). 
First, we observe, that at large $p$, the term $2pq$ dominates in the argument of the error function. We rewrite equation (\ref{eqn:deriv0040}) as
\begin{align}
  \sigma_p=\sqrt{\frac{\pi}{2}}\mathrm{Im}\,\Sigma(\mathbf{p},p^2)&=
  \sqrt{\frac{\pi}{2}}\frac{\omega_\mathrm{pl}^2}{4p}
  \int_{0}^{\infty}\frac{dq}{q\,\omega_\mathbf{q}}\Bigg\{ 
  \left[
  \text{Erf}\,\left( \frac{2pq}{\sqrt{2}\sigma_p} \right)-
  \text{Erf}\,\left( \frac{-2pq}{\sqrt{2}\sigma_p} \right)
  \right]\,n_\mathrm{B}(\omega_\mathbf{q})\,\exp(\omega_\mathbf{q}/T)\\
  &\quad-\left[
  \text{Erf}\,\left( \frac{2pq}{\sqrt{2}\sigma_p} \right)-
  \text{Erf}\,\left( \frac{-2pq}{\sqrt{2}\sigma_p} \right)
  \right]\,n_\mathrm{B}(-\omega_\mathbf{q})\,\exp(-\omega_\mathbf{q}/T)
  \Bigg\}\\
  &=
  \sqrt{\frac{\pi}{2}}\frac{\omega_\mathrm{pl}^2}{4p}
  \int_{0}^{\infty}\frac{dq}{q\,\omega_\mathbf{q}}
  2\text{Erf}\,\left( \frac{2pq}{\sqrt{2}\sigma_p} \right)\left[
  n_\mathrm{B}(\omega_\mathbf{q})\,\exp(\omega_\mathbf{q}/T)-n_\mathrm{B}(-\omega_\mathbf{q})\,\exp(-\omega_\mathbf{q}/T) \right]~,
  \label{eqn:deriv0050}
\end{align}

The integrand in equation~(\ref{eqn:deriv0050}) contains a steeply rising part at $q<-\sigma_p/p$ and a smoothly decaying part for at large
$q$, i.e. when $q\gg -\sigma_p/p$. Therefore, we 
separate the integral in equation  into two parts, one going from $q=0$ to $q=\bar q=-\sigma_p/p$ and the other from $\bar q$ to
infinity.
In the first part of the integral, the values for $q$ are so small, that we can replace the plasmon dispersion by the plasma frequency
$\omega_\mathrm{pl}$. In the second term, the argument of the error function is large and the error function can be replaced by its asymptotic
value at infinity, $\lim_{x\to\infty}\text{Erf}(x)=1$.
This leads to
\begin{multline}
  \sigma_p=\sqrt{\frac{\pi}{2}}\frac{\omega_\mathrm{pl}^2}{4p}\Bigg\{
  \int_{0}^{\bar q}\frac{dq}{q\,\omega_\mathrm{pl}}
  2\text{Erf}\,\left( \frac{2pq}{\sqrt{2}\sigma_p} \right)\left[
  n_\mathrm{B}(\omega_\mathrm{pl})\,\exp(\omega_\mathrm{pl}/T)-n_\mathrm{B}(-\omega_\mathrm{pl})\,\exp(-\omega_\mathrm{pl}/T) \right]\\
  +2\int_{\bar q}^{\infty}\frac{dq}{q\,\omega_\mathbf{q}}
  \left[
  n_\mathrm{B}(\omega_\mathbf{q})\,\exp(\omega_\mathbf{q}/T)-n_\mathrm{B}(-\omega_\mathbf{q})\,\exp(-\omega_\mathbf{q}/T) \right]
  \Bigg\}
  ~,
  \label{eqn:deriv0060}
\end{multline}
Finally, we expand last term in powers of $\omega_\mathbf{q}/T$, which is justified at low densities ($\omega_\mathbf{q}\propto \omega_\mathrm{pl}$),
and keep only the first order,
\begin{equation}
  n_\mathrm{B}(\omega_\mathbf{q})\,\exp(\omega_\mathbf{q}/T)-n_\mathrm{B}(-\omega_\mathbf{q})\,\exp(-\omega_\mathbf{q}/T)=\frac{2T}{\omega_\mathbf{q}}+\mathcal{O}(\omega_\mathbf{q})^{-3}~.
  \label{eqn:expansionBoseExp}
\end{equation}
We obtain
\begin{equation}
  \sigma_p=\sqrt{\frac{\pi}{2}}\frac{\omega_\mathrm{pl}^2}{4p}\Bigg\{
  \int_{0}^{\bar q}\frac{dq}{q\,\omega_\mathrm{pl}}
  2\text{Erf}\,\left( \frac{2pq}{\sqrt{2}\sigma_p} \right)\left[
  n_\mathrm{B}(\omega_\mathrm{pl})\,\exp(\omega_\mathrm{pl}/T)-n_\mathrm{B}(-\omega_\mathrm{pl})\,\exp(-\omega_\mathrm{pl}/T) \right]\\
  +4 T\int_{\bar q}^{\infty}\frac{dq}{q\,\omega_\mathbf{q}^2}\Bigg\}~.
\end{equation}
Both integrals can be performed analytically,
\begin{equation}
  \begin{split}
    \int_{0}^{\bar q}\frac{dq}{q}\text{Erf}\,\left( \frac{2pq}{\sqrt{2}\sigma_p} \right)=-2\sqrt{\frac{2}{\pi}}\frac{p\bar q}{\sigma_p}{}_2F_2(1/2,1/2;3/2,3/2;-2p^2 {\bar
  q}^2/\sigma_p^2)\\
  =-2\sqrt{\frac{2}{\pi}}{}_2F_2(1/2,1/2;3/2,3/2;-2)=-1.3357~,\\
  \int_{\bar q}^{\infty}\frac{dq}{q\,\omega_\mathrm{pl}^2(1+q^2/\kappa^2)}=\frac{1}{2} \ln(1 + \kappa^2/{\bar q}^2)=\frac{1}{2} \ln(1 +
  \kappa^2p^2/\sigma_p^2)~,
  \end{split}
  \label{eqn:integrals}
\end{equation}
where $\bar q=-\sigma_p/p$ was used.
Note that in the second integral, the $q^4$ term in the plasmon dispersion (\ref{eqn:GrossBohm}) is omitted. ${}_2F_2(a_1,a_2;b_1,b_2;z)$ is the
generalized hypergeometric function \cite{abra}.

We arrive at the equation
\begin{equation}
  \sigma_p=-1.3357\sqrt{\frac{\pi}{2}}\frac{\omega_\mathrm{pl}^2}{2p}\left[
  n_\mathrm{B}(\omega_\mathrm{pl})\,\exp(\omega_\mathrm{pl}/T)-n_\mathrm{B}(-\omega_\mathrm{pl})\,\exp(-\omega_\mathrm{pl}/T) \right]
  -\sqrt{\frac{\pi}{2}}\frac{T}{2p}\ln(1+\kappa^2p^2/\sigma_p^2)~.
  \label{eqn:deriv0070}
\end{equation}
At large $p$, the term $\kappa^2p^2/\sigma_p$ dominates the argument of the
logarithm, i.e. we can write $\ln(1+\kappa^2p^2/\sigma_p^2)\simeq \ln(\kappa^2p^2/\sigma_p^2)$.
Then, we arrive at equation (\ref{eqn:deriv0040}), given in section \ref{sec:analyticexpression}.

\section{Pade approximation\label{app:Pade}}
From the knowledge of the behaviour of $\sigma_p$ in the limits $p\to0$ and $p\to \infty$, a two point Pad\'e interpolation formula can be constructed.
For the value of the quasi-particle damping width at $p=0$ we take the expression
\begin{equation}
  \sigma_0=-\frac{\pi}{2}\sqrt{\kappa T}~,
  \label{eqn:sigma_p0}
\end{equation}
which is the exact solution of the self-consistent Born approximation
\cite{Fortmann_JPhysA41_445501_2008}. 

The Pad\'e interpolation formula is constructed in the following way:
We make the ansatz
\begin{equation}
  \sigma_p^\text{Pad\'e}=\frac{a_0+a_1 p}{1+b_1p+b_2p^2}\tilde\varphi(p)~,
  \label{appeqn:sigma_pade_ansatz}
\end{equation}
where the function $\tilde \varphi(p)$ contains the logarithmic terms present in the behaviour of $\sigma_p$ at large $p$, c.f. equation
(\ref{eqn:finalsolution}).
\begin{align}
  \tilde\varphi(p)&=\Bigg[\tilde\xi-\ln \tilde\xi +\frac{\ln \tilde\xi}{\tilde\xi}-\frac{\ln \tilde\xi}{\tilde\xi^2}+\frac{\ln \tilde\xi}{\tilde\xi}-\frac{3\ln^2
  \tilde\xi}{2\tilde\xi^3}+\frac{\ln^2\tilde\xi}{2\tilde\xi^2}+\frac{\ln^3\tilde\xi}{3\tilde\xi^3}\Bigg]~,\\
  \tilde \xi&=\ln(e+\sqrt{\frac{2}{\pi}}\kappa p^2\exp(A/T)/T)~.
  \label{appeqn:tilde}
\end{align}
The coefficients $a_0,a_1,b_1,b_2$ are determined by power expansion at $p=0$ and
$p\to \infty$, 
\begin{align}
  \lim_{p\to 0}\sigma_p^\text{Pad\'e}&=a_0+(a_1-a_0b_1)p+\mathcal{O}(p^2)~,\\
  \lim_{p\to\infty}\sigma_p^\text{Pad\'e}&=\left[\frac{a_1}{b_2 p}+\frac{a_0b_2-a_1b_1}{b_2^2 p^2}+\mathcal{O}(p^{-3})\right]\tilde\varphi(p)~,
  \label{eqn:Pade_deriv010}
\end{align}
and comparison to the behaviour of $\sigma_{p}$ in these limiting cases, e.g. equation (\ref{eqn:finalsolution}) for large $p$ and equation
(\ref{eqn:sigma_p0}) for $p\to 0$. Setting the slope of $\sigma_p$ at $p=0$ to zero, 
as well as the coefficient in front of the $p^{-2}$-term of the
asymptotic expansion, we arrive at the following equations for the coefficients of the interpolation formula,
\begin{xalignat}{2}{}
  a_0&=-\frac{\pi}{2}\sqrt{\kappa T}~,	&	a_1-a_0b_q&=0~,\\
  a_1&=-T\sqrt{\frac{\pi}{2}} b_2~,	&	a_0 b_2-a_1b_1&=0~.
\end{xalignat}
The solution reads
\begin{xalignat}{2}
  a_0&=-\frac{\pi}{2}\sqrt{\kappa T}~,	&	a_1&=-\kappa\left( \frac{\pi}{2} \right)^{3/2}~,\\
  b_1&=\sqrt{\frac{\pi\kappa}{2T}}~,	&	b_2&=\frac{\pi\kappa}{2T}~,
\end{xalignat}
which is given as equation (\ref{eqn:sigma_p_interpolation}) in the main text, section (\ref{sec:sf-se}).
 
\end{appendix}

%\bibliography{/home/roepke/carsten/latex/bibtex/citebase/all}

\begin{thebibliography}{46}
\expandafter\ifx\csname natexlab\endcsname\relax\def\natexlab#1{#1}\fi
\expandafter\ifx\csname bibnamefont\endcsname\relax
  \def\bibnamefont#1{#1}\fi
\expandafter\ifx\csname bibfnamefont\endcsname\relax
  \def\bibfnamefont#1{#1}\fi
\expandafter\ifx\csname citenamefont\endcsname\relax
  \def\citenamefont#1{#1}\fi
\expandafter\ifx\csname url\endcsname\relax
  \def\url#1{\texttt{#1}}\fi
\expandafter\ifx\csname urlprefix\endcsname\relax\def\urlprefix{URL }\fi
\providecommand{\bibinfo}[2]{#2}
\providecommand{\eprint}[2][]{\url{#2}}

\bibitem[{\citenamefont{Kadanoff and Baym}(1962)}]{KadanoffBaym:Book}
\bibinfo{author}{\bibfnamefont{L.~P.} \bibnamefont{Kadanoff}} \bibnamefont{and}
  \bibinfo{author}{\bibfnamefont{G.}~\bibnamefont{Baym}},
  \emph{\bibinfo{title}{Quantum Statistical Mechanics}}
  (\bibinfo{publisher}{W.A. Benjamin Inc.}, \bibinfo{address}{New York},
  \bibinfo{year}{1962}).

\bibitem[{\citenamefont{Mahan}(1981)}]{Mahan:Book}
\bibinfo{author}{\bibfnamefont{G.~D.} \bibnamefont{Mahan}},
  \emph{\bibinfo{title}{Many-Particle Physics}} (\bibinfo{publisher}{Plenum
  Press}, \bibinfo{address}{New York and London}, \bibinfo{year}{1981}),
  \bibinfo{edition}{2nd} ed.

\bibitem[{\citenamefont{Fetter and Walecka}(1971)}]{FetterWalecka_1971}
\bibinfo{author}{\bibfnamefont{A.~L.} \bibnamefont{Fetter}} \bibnamefont{and}
  \bibinfo{author}{\bibfnamefont{J.~D.} \bibnamefont{Walecka}},
  \emph{\bibinfo{title}{Quantum Theory of Many-Particle Systems}}
  (\bibinfo{publisher}{McGraw-Hill}, \bibinfo{address}{New York},
  \bibinfo{year}{1971}).

\bibitem[{\citenamefont{H{\"o}ll et~al.}(2006)\citenamefont{H{\"o}ll, Roberts,
  and Wright}}]{Hoell:2006AIPC..857...46H}
\bibinfo{author}{\bibfnamefont{A.}~\bibnamefont{H{\"o}ll}},
  \bibinfo{author}{\bibfnamefont{C.~D.} \bibnamefont{Roberts}},
  \bibnamefont{and} \bibinfo{author}{\bibfnamefont{S.~V.}
  \bibnamefont{Wright}}, in \emph{\bibinfo{booktitle}{Particles and Fields: X
  Mexican Workshop}}, edited by \bibinfo{editor}{\bibfnamefont{M.~A.}
  \bibnamefont{P{\'e}rez}},
  \bibinfo{editor}{\bibfnamefont{L.}~\bibnamefont{Urrutia}}, \bibnamefont{and}
  \bibinfo{editor}{\bibfnamefont{L.}~\bibnamefont{Villaseqor}}
  (\bibinfo{year}{2006}), vol. \bibinfo{volume}{857} of
  \emph{\bibinfo{series}{American Institute of Physics Conference Series}}, pp.
  \bibinfo{pages}{46--61}, \eprint{http://dx.doi.org/10.1063/1.2359242}.

\bibitem[{\citenamefont{Wierling}(2002)}]{Wierling_AdvPlasPhysRes_2002}
\bibinfo{author}{\bibfnamefont{A.}~\bibnamefont{Wierling}}, in
  \emph{\bibinfo{booktitle}{Advances in Plasma Physics Research}}, edited by
  \bibinfo{editor}{\bibfnamefont{F.}~\bibnamefont{Gerard}}
  (\bibinfo{publisher}{Nova Science}, \bibinfo{address}{New York},
  \bibinfo{year}{2002}), p. \bibinfo{pages}{127}.

\bibitem[{\citenamefont{Reinholz}(2005)}]{Reinholz:AnalDePhys06}
\bibinfo{author}{\bibfnamefont{H.}~\bibnamefont{Reinholz}},
  \bibinfo{journal}{Ann. Phys. Fr.} \textbf{\bibinfo{volume}{30}},
  \bibinfo{pages}{1} (\bibinfo{year}{2005}),
  \eprint{http://dx.doi.org/10.1051/anphys:2006004}.

\bibitem[{\citenamefont{Fortmann et~al.}(2007)\citenamefont{Fortmann,
  R{\"{o}}pke, and Wierling}}]{Fortmann:CPP47_2007}
\bibinfo{author}{\bibfnamefont{C.}~\bibnamefont{Fortmann}},
  \bibinfo{author}{\bibfnamefont{G.}~\bibnamefont{R{\"{o}}pke}},
  \bibnamefont{and} \bibinfo{author}{\bibfnamefont{A.}~\bibnamefont{Wierling}},
  \bibinfo{journal}{Contrib. Plasma Phys.} \textbf{\bibinfo{volume}{47}},
  \bibinfo{pages}{297} (\bibinfo{year}{2007}),
  \eprint{http://dx.doi.org/10.1002/ctpp.200710040}.

\bibitem[{\citenamefont{Reinholz et~al.}(2004)\citenamefont{Reinholz, Morozov,
  R{\"o}pke, and Millat}}]{Reinholz_PRE.69.066412_2008}
\bibinfo{author}{\bibfnamefont{H.}~\bibnamefont{Reinholz}},
  \bibinfo{author}{\bibfnamefont{I.}~\bibnamefont{Morozov}},
  \bibinfo{author}{\bibfnamefont{G.}~\bibnamefont{R{\"o}pke}},
  \bibnamefont{and} \bibinfo{author}{\bibfnamefont{T.}~\bibnamefont{Millat}},
  \bibinfo{journal}{Phys. Rev. E} \textbf{\bibinfo{volume}{69}},
  \bibinfo{pages}{066412} (\bibinfo{year}{2004}),
  \eprint{http://dx.doi.org/10.1103/PhysRevE.69.066412}.

\bibitem[{\citenamefont{Gericke et~al.}(1996)\citenamefont{Gericke, Schlanges,
  and Kraeft}}]{Gericke:PhysLettA222_241_1996}
\bibinfo{author}{\bibfnamefont{D.~O.} \bibnamefont{Gericke}},
  \bibinfo{author}{\bibfnamefont{M.}~\bibnamefont{Schlanges}},
  \bibnamefont{and} \bibinfo{author}{\bibfnamefont{W.-D.}
  \bibnamefont{Kraeft}}, \bibinfo{journal}{Phys. Lett. A}
  \textbf{\bibinfo{volume}{222}}, \bibinfo{pages}{5} (\bibinfo{year}{1996}),
  \eprint{http://dx.doi.org/10.1016/0375-9601(96)00654-8}.

\bibitem[{\citenamefont{Kraeft and
  Strege}(1988)}]{KraeftStrege_PhysicaA149_313_1988}
\bibinfo{author}{\bibfnamefont{W.-D.} \bibnamefont{Kraeft}} \bibnamefont{and}
  \bibinfo{author}{\bibfnamefont{B.}~\bibnamefont{Strege}},
  \bibinfo{journal}{Physica A} \textbf{\bibinfo{volume}{149}},
  \bibinfo{pages}{313} (\bibinfo{year}{1988}),
  \eprint{http://dx.doi.org/10.1016/0378-4371(88)90222-1}.

\bibitem[{\citenamefont{Vorberger et~al.}(2004)\citenamefont{Vorberger,
  Schlanges, and Kraeft}}]{Vorberger_PRE69_046407_2004}
\bibinfo{author}{\bibfnamefont{J.}~\bibnamefont{Vorberger}},
  \bibinfo{author}{\bibfnamefont{M.}~\bibnamefont{Schlanges}},
  \bibnamefont{and} \bibinfo{author}{\bibfnamefont{W.-D.}
  \bibnamefont{Kraeft}}, \bibinfo{journal}{Phys. Rev. E}
  \textbf{\bibinfo{volume}{69}}, \bibinfo{pages}{046407}
  (\bibinfo{year}{2004}),
  \eprint{http://dx.doi.org/10.1103/PhysRevE.69.046407}.

\bibitem[{\citenamefont{Kraeft et~al.}(1986)\citenamefont{Kraeft, Kremp,
  Ebeling, and R{\"{o}}pke}}]{KraeftKrempEbelingRoepke1986}
\bibinfo{author}{\bibfnamefont{W.~D.} \bibnamefont{Kraeft}},
  \bibinfo{author}{\bibfnamefont{D.}~\bibnamefont{Kremp}},
  \bibinfo{author}{\bibfnamefont{W.}~\bibnamefont{Ebeling}}, \bibnamefont{and}
  \bibinfo{author}{\bibfnamefont{G.}~\bibnamefont{R{\"{o}}pke}},
  \emph{\bibinfo{title}{Quantum Statistics of Charged Particle Systems}}
  (\bibinfo{publisher}{Akademie-Verlag}, \bibinfo{address}{Berlin},
  \bibinfo{year}{1986}).

\bibitem[{\citenamefont{Ebeling et~al.}(1972)\citenamefont{Ebeling, Kraeft, and
  Kremp}}]{EbelingKraeftKremp_1972}
\bibinfo{author}{\bibfnamefont{W.}~\bibnamefont{Ebeling}},
  \bibinfo{author}{\bibfnamefont{W.-D.} \bibnamefont{Kraeft}},
  \bibnamefont{and} \bibinfo{author}{\bibfnamefont{D.}~\bibnamefont{Kremp}},
  \emph{\bibinfo{title}{Theory of Bound States and Ionization Equilibrium in
  Plasmas and Solids}} (\bibinfo{publisher}{Akademie Verlag},
  \bibinfo{address}{Berlin}, \bibinfo{year}{1972}).

\bibitem[{\citenamefont{Dyson}(1949)}]{Dyson_PhysRev.75.1736}
\bibinfo{author}{\bibfnamefont{F.~J.} \bibnamefont{Dyson}},
  \bibinfo{journal}{Phys. Rev.} \textbf{\bibinfo{volume}{75}},
  \bibinfo{pages}{1736} (\bibinfo{year}{1949}),
  \eprint{http://dx.doi.org/10.1103/PhysRev.75.1736}.

\bibitem[{\citenamefont{Schwinger}(1951{\natexlab{a}})}]{Schwinger_PNAS37_452_%
1951}
\bibinfo{author}{\bibfnamefont{J.}~\bibnamefont{Schwinger}},
  \bibinfo{journal}{Proc. Nat. Acad. Sci.} \textbf{\bibinfo{volume}{37}},
  \bibinfo{pages}{451} (\bibinfo{year}{1951}{\natexlab{a}}),
  \eprint{http://www.pnas.org/content/37/7/452.full.pdf}.

\bibitem[{\citenamefont{Schwinger}(1951{\natexlab{b}})}]{Schwinger_PNAS37_455_%
1951}
\bibinfo{author}{\bibfnamefont{J.}~\bibnamefont{Schwinger}},
  \bibinfo{journal}{Proc. Nat. Acad. Sci.} \textbf{\bibinfo{volume}{37}},
  \bibinfo{pages}{455} (\bibinfo{year}{1951}{\natexlab{b}}),
  \eprint{http://www.pnas.org/content/37/7/455.full.pdf}.

\bibitem[{\citenamefont{Roberts and
  Schmidt}(2000)}]{RobertsSchmidt_ProgNucPartPhys45_2000}
\bibinfo{author}{\bibfnamefont{C.~D.} \bibnamefont{Roberts}} \bibnamefont{and}
  \bibinfo{author}{\bibfnamefont{S.~M.} \bibnamefont{Schmidt}},
  \bibinfo{journal}{Prog. Part. Nucl. Phys.} \textbf{\bibinfo{volume}{45}},
  \bibinfo{pages}{S1} (\bibinfo{year}{2000}),
  \eprint{http://dx.doi.org/10.1016/S0146-6410(00)90011-5}.

\bibitem[{\citenamefont{Hedin}(1965)}]{Hedin:PhysRev139_1965}
\bibinfo{author}{\bibfnamefont{L.}~\bibnamefont{Hedin}},
  \bibinfo{journal}{Phys. Rev.} \textbf{\bibinfo{volume}{139}},
  \bibinfo{pages}{796} (\bibinfo{year}{1965}),
  \eprint{http://dx.doi.org/10.1103/PhysRev.139.A796}.

\bibitem[{\citenamefont{von Barth et~al.}(2005)\citenamefont{von Barth, Dahlen,
  van Leeuwen, and Stefanucci}}]{Barth_PRB72_235109}
\bibinfo{author}{\bibfnamefont{U.}~\bibnamefont{von Barth}},
  \bibinfo{author}{\bibfnamefont{N.~E.} \bibnamefont{Dahlen}},
  \bibinfo{author}{\bibfnamefont{R.}~\bibnamefont{van Leeuwen}},
  \bibnamefont{and}
  \bibinfo{author}{\bibfnamefont{G.}~\bibnamefont{Stefanucci}},
  \bibinfo{journal}{Phys. Rev. B} \textbf{\bibinfo{volume}{72}},
  \bibinfo{eid}{235109} (\bibinfo{year}{2005}),
  \eprint{http://dx.doi.org/10.1103/PhysRevB.72.235109}.

\bibitem[{\citenamefont{Baym and
  Kadanoff}(1961)}]{BaymKadanoff_PhysRev.124.287}
\bibinfo{author}{\bibfnamefont{G.}~\bibnamefont{Baym}} \bibnamefont{and}
  \bibinfo{author}{\bibfnamefont{L.~P.} \bibnamefont{Kadanoff}},
  \bibinfo{journal}{Phys. Rev.} \textbf{\bibinfo{volume}{124}},
  \bibinfo{pages}{287} (\bibinfo{year}{1961}),
  \eprint{http://dx.doi.org/10.1103/PhysRev.124.287}.

\bibitem[{\citenamefont{Baym}(1962)}]{Baym_PhysRev.127.1391}
\bibinfo{author}{\bibfnamefont{G.}~\bibnamefont{Baym}}, \bibinfo{journal}{Phys.
  Rev.} \textbf{\bibinfo{volume}{127}}, \bibinfo{pages}{1391}
  (\bibinfo{year}{1962}), \eprint{http://dx.doi.org/10.1103/PhysRev.127.1391}.

\bibitem[{\citenamefont{Aryasetiawan and
  Gunnarsson}(1998)}]{Aryasetiawan:RepProgPhys61_1998}
\bibinfo{author}{\bibfnamefont{F.}~\bibnamefont{Aryasetiawan}}
  \bibnamefont{and}
  \bibinfo{author}{\bibfnamefont{O.}~\bibnamefont{Gunnarsson}},
  \bibinfo{journal}{Rep. Prog. Phys.} \textbf{\bibinfo{volume}{61}},
  \bibinfo{pages}{237} (\bibinfo{year}{1998}),
  \eprint{http://dx.doi.org10.1088/0034-4885/61/3/002}.

\bibitem[{\citenamefont{Onida et~al.}(2002)\citenamefont{Onida, Reining, and
  Rubio}}]{Onida:RevModPhys.74.601}
\bibinfo{author}{\bibfnamefont{G.}~\bibnamefont{Onida}},
  \bibinfo{author}{\bibfnamefont{L.}~\bibnamefont{Reining}}, \bibnamefont{and}
  \bibinfo{author}{\bibfnamefont{A.}~\bibnamefont{Rubio}},
  \bibinfo{journal}{Rev. Mod. Phys.} \textbf{\bibinfo{volume}{74}},
  \bibinfo{pages}{601} (\bibinfo{year}{2002}),
  \eprint{http://dx.doi.org/10.1103/RevModPhys.74.601}.

\bibitem[{\citenamefont{Mahan}(1994)}]{Mahan:CommCondMatPhys16_1994}
\bibinfo{author}{\bibfnamefont{G.~D.} \bibnamefont{Mahan}},
  \bibinfo{journal}{Comments Condens. Mat. Phys.}
  \textbf{\bibinfo{volume}{16}}, \bibinfo{pages}{333} (\bibinfo{year}{1994}).

\bibitem[{\citenamefont{van Hees and
  Knoll}(2001)}]{vHeesKnoll_PhysRevD.65.025010}
\bibinfo{author}{\bibfnamefont{H.}~\bibnamefont{van Hees}} \bibnamefont{and}
  \bibinfo{author}{\bibfnamefont{J.}~\bibnamefont{Knoll}},
  \bibinfo{journal}{Phys. Rev. D} \textbf{\bibinfo{volume}{65}},
  \bibinfo{pages}{025010} (\bibinfo{year}{2001}),
  \eprint{http://dx.doi.org/10.1103/PhysRevD.65.025010}.

\bibitem[{\citenamefont{van Hees and
  Knoll}(2002{\natexlab{a}})}]{PhysRevD.65.105005}
\bibinfo{author}{\bibfnamefont{H.}~\bibnamefont{van Hees}} \bibnamefont{and}
  \bibinfo{author}{\bibfnamefont{J.}~\bibnamefont{Knoll}},
  \bibinfo{journal}{Phys. Rev. D} \textbf{\bibinfo{volume}{65}},
  \bibinfo{pages}{105005} (\bibinfo{year}{2002}{\natexlab{a}}),
  \eprint{http://dx.doi.org/10.1103/PhysRevD.65.105005}.

\bibitem[{\citenamefont{van Hees and
  Knoll}(2002{\natexlab{b}})}]{PhysRevD.66.025028}
\bibinfo{author}{\bibfnamefont{H.}~\bibnamefont{van Hees}} \bibnamefont{and}
  \bibinfo{author}{\bibfnamefont{J.}~\bibnamefont{Knoll}},
  \bibinfo{journal}{Phys. Rev. D} \textbf{\bibinfo{volume}{66}},
  \bibinfo{pages}{025028} (\bibinfo{year}{2002}{\natexlab{b}}),
  \eprint{http://dx.doi.org/10.1103/PhysRevD.66.025028}.

\bibitem[{\citenamefont{Takada}(2001)}]{Takada:PhysRevLett.87.226402}
\bibinfo{author}{\bibfnamefont{Y.}~\bibnamefont{Takada}},
  \bibinfo{journal}{Phys. Rev. Lett.} \textbf{\bibinfo{volume}{87}},
  \bibinfo{pages}{226402} (\bibinfo{year}{2001}),
  \eprint{http://dx.doi.org/10.1103/PhysRevLett.87.226402}.

\bibitem[{\citenamefont{von Barth and Holm}(1996)}]{BarthHolm:PRB54_1996}
\bibinfo{author}{\bibfnamefont{U.}~\bibnamefont{von Barth}} \bibnamefont{and}
  \bibinfo{author}{\bibfnamefont{B.}~\bibnamefont{Holm}},
  \bibinfo{journal}{Phys. Rev. B} \textbf{\bibinfo{volume}{54}},
  \bibinfo{pages}{8411} (\bibinfo{year}{1996}),
  \eprint{http://dx.doi.org/10.1103/PhysRevB.54.8411}.

\bibitem[{\citenamefont{Holm and von Barth}(1998)}]{Holm:prb57_98}
\bibinfo{author}{\bibfnamefont{B.}~\bibnamefont{Holm}} \bibnamefont{and}
  \bibinfo{author}{\bibfnamefont{U.}~\bibnamefont{von Barth}},
  \bibinfo{journal}{Phys. Rev. B} \textbf{\bibinfo{volume}{57}},
  \bibinfo{pages}{2108} (\bibinfo{year}{1998}),
  \eprint{http://dx.doi.org/0.1103/PhysRevB.57.2108}.

\bibitem[{\citenamefont{Fehr and Kraeft}(1995)}]{FehrKraeft_CPP35_463_1995}
\bibinfo{author}{\bibfnamefont{R.}~\bibnamefont{Fehr}} \bibnamefont{and}
  \bibinfo{author}{\bibfnamefont{W.-D.} \bibnamefont{Kraeft}},
  \bibinfo{journal}{Contrib. Plasma Phys.} \textbf{\bibinfo{volume}{35}},
  \bibinfo{pages}{463} (\bibinfo{year}{1995}),
  \eprint{http://dx.doi.org/0.1002/ctpp.2150350602}.

\bibitem[{\citenamefont{Wierling and R{\"o}pke}(1998)}]{Wierling:CPP38_1998}
\bibinfo{author}{\bibfnamefont{A.}~\bibnamefont{Wierling}} \bibnamefont{and}
  \bibinfo{author}{\bibfnamefont{G.}~\bibnamefont{R{\"o}pke}},
  \bibinfo{journal}{Contrib. Plasma Phys.} \textbf{\bibinfo{volume}{38}},
  \bibinfo{pages}{513} (\bibinfo{year}{1998}),
  \eprint{http://dx.doi.org/10.1002/ctpp.2150380405}.

\bibitem[{\citenamefont{{Schepe} et~al.}(1998)\citenamefont{{Schepe},
  {Schmielau}, {Tamme}, and {Henneberger}}}]{Schepe-Schmielau:PSSb206_1998}
\bibinfo{author}{\bibfnamefont{R.}~\bibnamefont{{Schepe}}},
  \bibinfo{author}{\bibfnamefont{T.}~\bibnamefont{{Schmielau}}},
  \bibinfo{author}{\bibfnamefont{D.}~\bibnamefont{{Tamme}}}, \bibnamefont{and}
  \bibinfo{author}{\bibfnamefont{K.}~\bibnamefont{{Henneberger}}},
  \bibinfo{journal}{Phys. Status Solidi B} \textbf{\bibinfo{volume}{206}},
  \bibinfo{pages}{273} (\bibinfo{year}{1998}),
  \eprint{http://dx.doi.org/10.1002/(SICI)1521-3951(199803)206:1<273::AID-PSSB%
273>3.0.CO;2-T}.

\bibitem[{\citenamefont{Fennel and
  Wilfer}(1974)}]{FennelWilfer_AnnPhysL32_265_1974}
\bibinfo{author}{\bibfnamefont{W.}~\bibnamefont{Fennel}} \bibnamefont{and}
  \bibinfo{author}{\bibfnamefont{H.~P.} \bibnamefont{Wilfer}},
  \bibinfo{journal}{Ann. Phys. Lpz.} \textbf{\bibinfo{volume}{32}},
  \bibinfo{pages}{265} (\bibinfo{year}{1974}),
  \eprint{http://dx.doi.org/10.1002/andp.19754870406}.

\bibitem[{\citenamefont{Fortmann}(2008)}]{Fortmann_JPhysA41_445501_2008}
\bibinfo{author}{\bibfnamefont{C.}~\bibnamefont{Fortmann}},
  \bibinfo{journal}{J. Phys. A: Math. Theor.} \textbf{\bibinfo{volume}{41}},
  \bibinfo{pages}{445501} (\bibinfo{year}{2008}),
  \eprint{http://arxiv.org/abs/0805.4674}.

\bibitem[{\citenamefont{Bahcall et~al.}(1995)\citenamefont{Bahcall,
  Pinsonneault, and Wasserburg}}]{Bahcall:RMP67.781.1995}
\bibinfo{author}{\bibfnamefont{J.~N.} \bibnamefont{Bahcall}},
  \bibinfo{author}{\bibfnamefont{M.~H.} \bibnamefont{Pinsonneault}},
  \bibnamefont{and} \bibinfo{author}{\bibfnamefont{G.~J.}
  \bibnamefont{Wasserburg}}, \bibinfo{journal}{Rev. Mod. Phys.}
  \textbf{\bibinfo{volume}{67}}, \bibinfo{pages}{781} (\bibinfo{year}{1995}),
  \eprint{http://dx.doi.org/10.1103/RevModPhys.67.781}.

\bibitem[{\citenamefont{Lundquist}(1967)}]{Lundquist_PhysKondMat6_193_1967}
\bibinfo{author}{\bibfnamefont{B.~I.} \bibnamefont{Lundquist}},
  \bibinfo{journal}{Phys. Kond. Mater.} \textbf{\bibinfo{volume}{6}},
  \bibinfo{pages}{206} (\bibinfo{year}{1967}).

\bibitem[{\citenamefont{Bohm and Gross}(1949)}]{GrossBohm_PhysRev.75.1851}
\bibinfo{author}{\bibfnamefont{D.}~\bibnamefont{Bohm}} \bibnamefont{and}
  \bibinfo{author}{\bibfnamefont{E.~P.} \bibnamefont{Gross}},
  \bibinfo{journal}{Phys. Rev.} \textbf{\bibinfo{volume}{75}},
  \bibinfo{pages}{1851} (\bibinfo{year}{1949}),
  \eprint{http://dx.doi.org/10.1103/PhysRev.75.1851}.

\bibitem[{\citenamefont{Thiele et~al.}(2008)\citenamefont{Thiele, Bornath,
  Fortmann, H\"{o}ll, Redmer, Reinholz, R\"{o}pke, Wierling, Glenzer, and
  Gregori}}]{Thiele_PRE_accepted}
\bibinfo{author}{\bibfnamefont{R.}~\bibnamefont{Thiele}},
  \bibinfo{author}{\bibfnamefont{T.}~\bibnamefont{Bornath}},
  \bibinfo{author}{\bibfnamefont{C.}~\bibnamefont{Fortmann}},
  \bibinfo{author}{\bibfnamefont{A.}~\bibnamefont{H\"{o}ll}},
  \bibinfo{author}{\bibfnamefont{R.}~\bibnamefont{Redmer}},
  \bibinfo{author}{\bibfnamefont{H.}~\bibnamefont{Reinholz}},
  \bibinfo{author}{\bibfnamefont{G.}~\bibnamefont{R\"{o}pke}},
  \bibinfo{author}{\bibfnamefont{A.}~\bibnamefont{Wierling}},
  \bibinfo{author}{\bibfnamefont{S.~H.} \bibnamefont{Glenzer}},
  \bibnamefont{and} \bibinfo{author}{\bibfnamefont{G.}~\bibnamefont{Gregori}},
  \bibinfo{journal}{Phys. Rev. E} \textbf{\bibinfo{volume}{78}},
  \bibinfo{eid}{026411} (\bibinfo{year}{2008}),
  \eprint{http://dx.doi.org/10.1103/PhysRevE.78.026411}.

\bibitem[{\citenamefont{Knoll and Voskresensky}(1996)}]{knol:annals96}
\bibinfo{author}{\bibfnamefont{J.}~\bibnamefont{Knoll}} \bibnamefont{and}
  \bibinfo{author}{\bibfnamefont{D.}~\bibnamefont{Voskresensky}},
  \bibinfo{journal}{Ann. Phys. NY} \textbf{\bibinfo{volume}{249}},
  \bibinfo{pages}{532} (\bibinfo{year}{1996}),
  \eprint{http://dx.doi.org/10.1006/aphy.1996.0082}.

\bibitem[{\citenamefont{Migdal}(1956)}]{migd:physrev56}
\bibinfo{author}{\bibfnamefont{A.~B.} \bibnamefont{Migdal}},
  \bibinfo{journal}{Phys. Rev.} \textbf{\bibinfo{volume}{103}},
  \bibinfo{pages}{1811} (\bibinfo{year}{1956}),
  \eprint{http://dx.doi.org/10.1103/PhysRev.103.1811}.

\bibitem[{\citenamefont{Klein}(1999)}]{klei:revmodphys99}
\bibinfo{author}{\bibfnamefont{S.}~\bibnamefont{Klein}}, \bibinfo{journal}{Rev.
  Mod. Phys.} \textbf{\bibinfo{volume}{71}}, \bibinfo{pages}{1501}
  (\bibinfo{year}{1999}),
  \eprint{http://dx.doi.org/10.1103/RevModPhys.71.1501}.

\bibitem[{\citenamefont{Hansen et~al.}(2003)\citenamefont{Hansen, Uggerh{\o}j,
  Biino, Ballestrero, Mangiarotti, Sona, Ketel, and Vilakazi}}]{hans:prl03}
\bibinfo{author}{\bibfnamefont{H.~D.} \bibnamefont{Hansen}},
  \bibinfo{author}{\bibfnamefont{U.~I.} \bibnamefont{Uggerh{\o}j}},
  \bibinfo{author}{\bibfnamefont{C.}~\bibnamefont{Biino}},
  \bibinfo{author}{\bibfnamefont{S.}~\bibnamefont{Ballestrero}},
  \bibinfo{author}{\bibfnamefont{A.}~\bibnamefont{Mangiarotti}},
  \bibinfo{author}{\bibfnamefont{P.}~\bibnamefont{Sona}},
  \bibinfo{author}{\bibfnamefont{T.~J.} \bibnamefont{Ketel}}, \bibnamefont{and}
  \bibinfo{author}{\bibfnamefont{Z.~Z.} \bibnamefont{Vilakazi}},
  \bibinfo{journal}{Phys. Rev. Lett.} \textbf{\bibinfo{volume}{91}},
  \bibinfo{pages}{014801} (\bibinfo{year}{2003}),
  \eprint{http://dx.doi.org/10.1103/PhysRevLett.91.014801}.

\bibitem[{\citenamefont{Hansen et~al.}(2004)\citenamefont{Hansen, Uggerh{\o}j,
  Biino, Ballestrero, Mangiarotti, Sona, Ketel, and Vilakazi}}]{hans:prd04}
\bibinfo{author}{\bibfnamefont{H.~D.} \bibnamefont{Hansen}},
  \bibinfo{author}{\bibfnamefont{U.~I.} \bibnamefont{Uggerh{\o}j}},
  \bibinfo{author}{\bibfnamefont{C.}~\bibnamefont{Biino}},
  \bibinfo{author}{\bibfnamefont{S.}~\bibnamefont{Ballestrero}},
  \bibinfo{author}{\bibfnamefont{A.}~\bibnamefont{Mangiarotti}},
  \bibinfo{author}{\bibfnamefont{P.}~\bibnamefont{Sona}},
  \bibinfo{author}{\bibfnamefont{T.~J.} \bibnamefont{Ketel}}, \bibnamefont{and}
  \bibinfo{author}{\bibfnamefont{Z.~Z.} \bibnamefont{Vilakazi}},
  \bibinfo{journal}{Phys. Rev. D} \textbf{\bibinfo{volume}{69}},
  \bibinfo{pages}{032001} (\bibinfo{year}{2004}),
  \eprint{http://dx.doi.org/10.1103/PhysRevD.69.032001}.

\bibitem[{\citenamefont{Fortmann et~al.}(2005)\citenamefont{Fortmann, Reinholz,
  R\"{o}pke, and Wierling}}]{Fortmann:CMT20_2007}
\bibinfo{author}{\bibfnamefont{C.}~\bibnamefont{Fortmann}},
  \bibinfo{author}{\bibfnamefont{H.}~\bibnamefont{Reinholz}},
  \bibinfo{author}{\bibfnamefont{G.}~\bibnamefont{R\"{o}pke}},
  \bibnamefont{and} \bibinfo{author}{\bibfnamefont{A.}~\bibnamefont{Wierling}},
  in \emph{\bibinfo{booktitle}{Condensed Matter Theory}}, edited by
  \bibinfo{editor}{\bibfnamefont{J.}~\bibnamefont{Clark}},
  \bibinfo{editor}{\bibfnamefont{R.}~\bibnamefont{Panoff}}, \bibnamefont{and}
  \bibinfo{editor}{\bibfnamefont{H.}~\bibnamefont{Li}}
  (\bibinfo{publisher}{Nova Science}, \bibinfo{address}{New York},
  \bibinfo{year}{2005}), vol.~\bibinfo{volume}{20}, p. \bibinfo{pages}{317},
  \urlprefix\url{http://arxiv.org/abs/physics/0502051}.

\bibitem[{\citenamefont{Abramowitz and Stegun}(1970)}]{abra}
\bibinfo{editor}{\bibfnamefont{M.}~\bibnamefont{Abramowitz}} \bibnamefont{and}
  \bibinfo{editor}{\bibfnamefont{A.}~\bibnamefont{Stegun}}, eds.,
  \emph{\bibinfo{title}{Handbook of Mathematical Functions with Formulas,
  Graphs and Mathematical Tables}} (\bibinfo{publisher}{Dover Publications},
  \bibinfo{address}{New York}, \bibinfo{year}{1970}), \bibinfo{edition}{9th}
  ed.

\end{thebibliography}

\end{document}